\documentclass[12pt, draftclsnofoot, journal, letterpaper, onecolumn]{IEEEtran}

\makeatletter
\def\ps@headings{%
\def\@oddhead{\mbox{}\scriptsize\rightmark \hfil \thepage}%
\def\@evenhead{\scriptsize\thepage \hfil \leftmark\mbox{}}%
\def\@oddfoot{}%
\def\@evenfoot{}}
\makeatother 
\IEEEoverridecommandlockouts
\usepackage{bbm}
\usepackage{amsfonts}
\usepackage[dvips]{graphicx}
\usepackage{times}
\usepackage{cite}
\usepackage{amsmath}
\usepackage{array}
\usepackage{amssymb}

\newcommand{\bs}{\boldsymbol}

\usepackage{stfloats}
\usepackage{graphicx}
\usepackage{footnote}
\usepackage{booktabs}
\usepackage{array}
\usepackage{algorithm}
\usepackage{subeqnarray}
\usepackage{cases}
\usepackage{threeparttable}
\usepackage{color}
\usepackage{hyperref}
\hypersetup{hidelinks} 
\usepackage{epstopdf}
\usepackage{algpseudocode}
\usepackage{bm}
\usepackage{multirow}
\usepackage[labelformat=simple,font=small]{subcaption}
\usepackage{adjustbox}

\usepackage{enumitem}

\usepackage{xcolor,etoolbox}

\usepackage{tikz}

\let\mybibitem\bibitem
\renewcommand{\bibitem}[1]{%
\ifstrequal{#1}{8778671}{\color{black}\mybibitem{#1}}
{\ifstrequal{#1}{wang2018preempt}{\color{black}\mybibitem{#1}}
{\ifstrequal{#1}{kavitha2018controlling}{\color{black}\mybibitem{#1}}
{\color{black}\mybibitem{#1}}}}%
}

\newtheorem{theorem}{\bf Theorem}

\newtheorem{lemma}{\bf Lemma}

\allowdisplaybreaks

\begin{document}

\title{ Performance Analysis of Age of Information in Ultra-Dense Internet of Things  (IoT) Systems with Noisy Channels}

\author{Bo~Zhou and Walid~Saad,~\IEEEmembership{Fellow,~IEEE}
\vspace{-1cm}
\thanks{
This work was supported by the Office of Naval Research (ONR) under MURI Grant N00014-19-1-2621.
A preliminary version of this work\cite{noisy_conf} was submitted for conference publication.

B.~Zhou and W.~Saad are with Wireless@VT, Bradley Department of Electrical and Computer Engineering, Virginia Tech, Blacksburg, VA 24061, USA.
Email: \{ecebo, walids\}@vt.edu.}
}

\maketitle

\begin{abstract}
In this paper, a dense Internet of Things (IoT) monitoring system is studied in which a large number of devices contend for transmitting timely status packets to their corresponding receivers over wireless noisy channels, using a carrier sense multiple access (CSMA) scheme.
When each device completes one transmission, due to possible transmission failure, two cases with and without transmission feedback to each device must be considered.
Particularly, for the case with no feedback,  the device uses policy (I): It will  go to an idle state and release the channel regardless of the outcome of the transmission.
For the case with perfect feedback, if the transmission succeeds, the device will go to an idle state, otherwise it uses either policy (W), i.e., it will go to a waiting state and re-contend for channel access; or it uses policy (S), i.e., it will stay at a service state and occupy this channel to attempt another transmission.
For those three policies, the closed-form expressions of the average
age of information (AoI) of each device are characterized under schemes with and without preemption in service. 
It is shown that, for each policy, the scheme with preemption in service always achieves a smaller average AoI, compared with the scheme without preemption. 
Then, a mean-field approximation approach with guaranteed accuracy is developed to analyze the asymptotic performance for the considered system with an infinite number of devices and the effects of the system parameters on the average AoI are characterized.
Simulation results show that the proposed mean-field approximation is accurate even for a small number of devices. The results also show that policy (S) achieves the smallest average AoI compared with policies (I) and (W), and the average AoI does not always decrease with the arrival rate for all three policies.
\end{abstract} 

\begin{IEEEkeywords}
Internet of things, status update, age of information, mean field, CSMA.
\end{IEEEkeywords}

 \section{Introduction}
Owing to the rapid developments of wireless communication technologies, the next-generation Internet of Things (IoT) will encompass a large
number of IoT devices that must send fresh information updates of various real-world physical processes to a myriad of time-critical  IoT applications\cite{8869705,9060999}.
To quantify the information freshness in these real-time IoT applications, the concept of \emph{age of information} (AoI) has emerged as a fundamental performance metric that is germane to various communication systems\cite{8778671,8938128,chaccour2020ruin,9149370,8469047,li2020age,zhang2020pricing,ornee2019sampling-arxiv,ferdowsi2020neural,feng2018age,9155514,li2019waiting,9184001}.
For practical ultra-dense IoT systems, due to the difficulty of performing coordinated channel access with a centralized unit and the requirements of time synchronization among all the devices\cite{zucchetto2017uncoordinated},
 it is of great importance to investigate the AoI performance under uncoordinated channel access.
 The key challenges in such analysis include the characterization of the complex temporal evolution of the AoI and the strong coupling among a large number of devices while accessing the channels. 

There have been several recent works on the analysis of the AoI under uncoordinated channel access, that can be classified into two broad groups based on the type their adopted. The first group in \cite{8006544,Chen2019,Chen2020,yang2020age,munari2020modern} considers ALOHA-like random access schemes, under which each device transmits its status packet to the receiver in each slot with a certain probability. 
Particularly, the authors in \cite{8006544} study the optimal attempt probability to minimize the average AoI in a multiaccess channel.
The works in \cite{Chen2019,Chen2020,yang2020age} analyze the average AoI for distributed randomized transmission polices.
The work in \cite{munari2020modern} considers an irregular repetition slotted ALOHA  protocol and derives expressions of the average AoI and the peak age violation probability.
Note that, in a real-world dense IoT, ALOHA-like random access schemes may lead to significant collisions, effectively rendering communication highly unreliable\cite{zucchetto2017uncoordinated}.
To reduce the transmission collisions, the second group of works in \cite{bedewy2019optimizing,9007478,8901143,zhou2020meanfield} considers carrier sense multiple access (CSMA) schemes that enable  the devices with carrier sensing capabilities so that they could sense the channel before updating the status packets.
Particularly, the work in \cite{bedewy2019optimizing} considers the use of sleep-wake scheduling to minimize the total average peak AoI while the work in \cite{9007478} considers the problem of minimizing the total average AoI by controlling the backoff times of the devices.
In \cite{8901143}, the authors analyze the worst-case AoI performance from the view of one device for a system in which all the other devices always have packets to send.
In \cite{zhou2020meanfield}, we studied the average AoI and the average peak AoI for a ultra-dense IoT and proposed a mean-field game in which each device optimizes its waiting rate so as to minimize the AoI performance. 

In the existing literature on the analysis of AoI of ultra-dense IoT systems with random packet arrivals under CSMA, e.g., \cite{9007478,8901143,zhou2020meanfield},  the transmission of each status packet is assumed to be always successful. 
In the presence of \emph{noisy channels}, the fundamental problem is to investigate how the possible failure of each transmission affects the AoI performance of each device.
This problem has not been considered in the prior art \cite{9007478,8901143,zhou2020meanfield}.
To address this problem, one must consider the effects of the existence of \emph{an ACK/NACK feedback channel}, through which each device will be immediately informed on whether or not its transmission is successful.
Particularly, when a device completes one transmission, if there is no feedback, then that device will release the channel regardless of the outcome of the transmission. In contrast, if there exists perfect feedback and the transmission fails, then the device may release the channel or still occupy it until its packet is delivered.
To date, the achievable AoI performance remains unknown  for ultra-dense IoT systems of noisy channels with and without feedback under CSMA.

The main contribution of this paper is, thus, a rigorous analytical characterization of the average AoI for an ultra-dense IoT system with random status packets arrivals under a CSMA-type access scheme, in presence of noisy channels.
We consider two cases: With and without feedback.
In particular, for each device, instead of transmitting a newly arriving status packet immediately,
 the device senses one channel to check whether it is busy and waits for a random period of time.
Thus, for each device, there are three possible states: Idle (\textit{I}), waiting (\textit{W}), and service  (\textit{S}).
When a device  in state \textit{S} completes one transmission,  for the case with no feedback, we consider policy (I) under which each device always moves to state \textit{I}.
Meanwhile, for the case with perfect feedback, we consider two policies: policy (W) and policy (S), under which, the device moves to state \textit{W} or stays at state \textit{S}, respectively, in case of transmission failure.

 Using tools from stochastic hybrid systems (SHS)\cite{8469047}, we derive the closed-form expressions of the average AoI of each device for policies (I), (W), and (S) under schemes with and without preemption in service.
 We show that, for each policy, the scheme with preemption in service always achieves better average AoI performance compared to the scheme without preemption.
Then,  we  analyze the asymptotic performance of the three policies for the considered IoT system  in the mean-field regime when the number of
devices grows large by using a mean-field approximation \cite{10.1145/3084454,10.1145/2964791.2901463,10.1145/2825236.2825241}. 
We also characterize the influence of the system parameters on the average AoI in the mean-field regime and show that, under policies (I) and (S), the systems with higher service rates, higher waiting rates, more communication resources, and better channel conditions can achieve a smaller average AoI.
Simulations validate our analytical results and show that the proposed mean-field approximation is very accurate even for a small number of devices.
Moreover, we show that policy (S) always achieves the smallest average AoI compared with policies (I) and (S). Meanwhile, our results also show that policy (W) only outperforms policy (I) when the channel utilization is small.
We also observe that the average AoI of each policy does not necessarily decrease with the arrival rate.
This contradicts the intuition that the average AoI should always decreases with the arrival rate for systems in which preemption in waiting/service can be allowed, for example, as seen in \cite{8469047} and \cite{9007478}.

The rest of this paper is organized as follows. 
Section~\ref{sec:systemmodel} presents the system model and the CSMA-type random access scheme.
In Section~\ref{sec:aoi-analysis}, we analyze the average AoI under the three policies with two  packet management schemes.
In Section~\ref{sec:mean-field}, we propose a mean-field framework and analyze the properties of the average AoI.
Section~\ref{sec:simulations} presents and analyzes numerical results.
Finally, conclusions are drawn in Section~\ref{sec:conclusion}.

\section{System Model}\label{sec:systemmodel}

\begin{figure}[!t]
\begin{centering}
\includegraphics[scale=0.8]{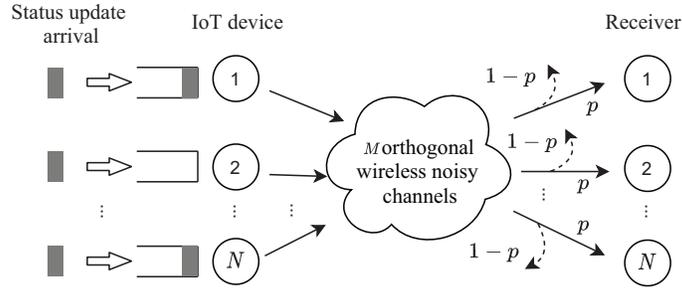}
\vspace{-0.1cm}
\caption{\small{Illustration of a real-time IoT monitoring system with $N$ pairs of IoT devices and their corresponding receivers, and $M$ orthogonal wireless noisy channels.}}\label{fig:system}
\end{centering}
\vspace{-0.35cm}
\end{figure}

As illustrated in Fig.~\ref{fig:system}, we consider a real-time ultra-dense IoT monitoring system composed of $N$ identical pairs of IoT devices (transmitters) and the corresponding receivers, and $M=N/\gamma$ identical orthogonal wireless noisy channels. 
Each IoT device monitors the associated underlying physical process and sends a real-time information status packet to its corresponding receiver.
We consider that the status packets of each each device's associated physical process randomly arrive according to a Poisson process with rate $\lambda$ and that the transmission time for each packet over each channel is exponentially distributed with the same mean $1/\mu$.
We assume that each device can occupy at most one channel in the transmission of one packet, as commonly done in prior art \cite{munari2020modern,Chen2020,yang2020age,8901143,9007478,bedewy2019optimizing,zhou2020meanfield}.
To avoid transmission collisions, we further assume that multiple devices with carrier sensing capabilities cannot concurrently transmit their packets over any given channel.

We consider that each channel between a given device and its corresponding receiver is noisy. In particular, upon transmission completion, each packet in service arrives successfully at the corresponding receiver with probability $p\in[0,1]$.
As done in \cite{feng2018age}, we consider two possible cases depending on whether or not there exists any transmission feedback to each device from the associated receiver.
For the case with no feedback, each IoT device would not know whether the packet is successfully delivered. Meanwhile, in the case with perfect feedback, each device will be immediately informed on whether or not its transmission is successful.

\subsection{CSMA-Type Random Channel Access}

We consider a CSMA-type random channel access scheme in which each device senses the channel before transmitting a status packet. 
Hence, for each device, there are three possible states: Idle (\textit{I}), waiting (\textit{W}), and service (\textit{S}).
In particular, if there is a new status packet arriving at a device in state \textit{I}, then prior to sending it immediately (i.e., going to state \textit{S}), the device will sense one of the channels to determine whether it is occupied and go to state \textit{W}.  If the channel is sensed to be busy, then the device remains silent; otherwise, the device backs off for a random period of time, that is exponentially distributed with rate $w$. During the waiting period, the device keeps sensing the channel to identify any conflicting transmissions. If any such transmission is found, the device will suspend its backoff timer and resume it when the channel is sensed idle. 
As is commonly done in the literature (e.g., \cite{10.1145/2825236.2825241,9007478} and \cite{5340575}), we adopt idealized CSMA assumptions, i.e., the channel sensing is instantaneous and there are no hidden nodes. 
Under these two assumptions and given  the continuous distributions of the backoff times, the probability of two devices starting their transmissions  simultaneously is zero\cite{5340575}. 

For each device, during the waiting period, any arriving  packet will replace the older one that is waiting at the device, because the receiver will not benefit from an outdated  packet.
During the service period, we consider two packet management schemes depending on whether or not the packet currently in service can be preempted. In particular, under a \emph{scheme with preemption in service (WP)}, the  packet currently in service will be preempted by a newly arrived one and then be discarded. Meanwhile, under a \emph{scheme without preemption (WOP)}, the arriving status packets will be discarded.

When a device in state \textit{S} completes the transmission of one status packet in service, we need to determine which state this device should move to, depending on whether this device receives feedback or not.
For the case with no feedback, the device  moves to state \textit{I} regardless of whether the transmission is successful or not. This policy is referred to as policy (I).
For the case with perfect feedback, we consider two policies: Policies (W) and  (S).  
Specifically, if the transmission is successful, under both policies, the device moves to state \textit{I}. In contrast, if the transmission fails, under policy (W), it moves back to state \textit{W} and waits for another channel access opportunity, or under policy (S), it stays at state \textit{S} and continuously occupies this channel until the transmission is successful.

\subsection{State Transitions of Each IoT Device}

Let $D_n(t)\in\mathcal{D}\triangleq\{\textit{I},\textit{W},\textit{S}\}$ be the state of device $n$ at time $t$.
Then, let $\bs{X}(t)\triangleq (X_d(t))_{d\in\mathcal{D}}$ be the empirical measure at time $t$, where $X_d(t) \triangleq \frac{1}{N}\sum_{n=1}^N \bs{1}\left(D_n(t)=d\right)$ represents the fraction of devices in state $d\in\mathcal{D}$ at $t$, with $\bs{1}(\cdot)$ being the indicator function.
Since all devices are exchangeable, we have 
$\mathbb{E}[X_d(t)] = \frac{1}{N}\sum_{n=1}^N \Pr[D_n(t)=d] = \Pr[D_n(t)=d]$ 
\cite{10.1145/3084454}.

Under policies (I), (W), and (S) with schemes WP and WOP, the process $\{D_n(t)\}$ of each device will be a continuous-time Markov chain (CTMC), with the following transitions:
\begin{itemize}
	\item  Under the three policies, when the device is in state \textit{I}, if there is a new status packet arrival, then the device will move to state \textit{W}, with rate $\lambda$.
	\item Under the three policies, when the device is in state \textit{W}, the probability of finding an idle channel is $1-\frac{1}{M}NX_{\textit{S}}(t) = 1-\gamma X_{\textit{S}}(t)$. Thus, the device transitions from state \textit{W} to state \textit{S} with rate $w(1-\gamma X_{\textit{S}}(t))$.
	\item When the device is in state \textit{S}, for scheme WOP, it is obvious that the service rate is $\mu$. For scheme WP, although a random number of  packets in service may be preempted, the service rate is also $\mu$. This is because the service period is memoryless in nature, and it is independent of the number of  packets in service that get preempted\cite{8469047}. Thus, for each one of the three policies, the CTMC under scheme WP is the same to the one under scheme WOP.
	Particularly, under policy (I), the device transitions from state \textit{S} to state \textit{I} with rate $\mu$. Under policy (W), the device moves from state \textit{S} to state \textit{I} with rate $\mu p$ and from state \textit{S} to state \textit{W} with rate $\mu(1-p)$. Under policy (S), the device transitions from state \textit{S} to state \textit{I} with rate $\mu p$.
\end{itemize}

\begin{figure}[!t]
\centering
\begin{minipage}[h]{.325\linewidth}{}
\centering
       \includegraphics[scale=1]{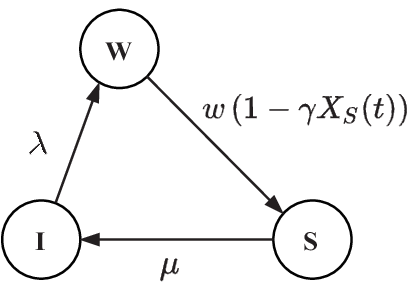}
\subcaption{Policy (I)}\label{fig:CTMC-policy-I}
\end{minipage}
\begin{minipage}[h]{.325\linewidth}
\centering
        \includegraphics[scale=1]{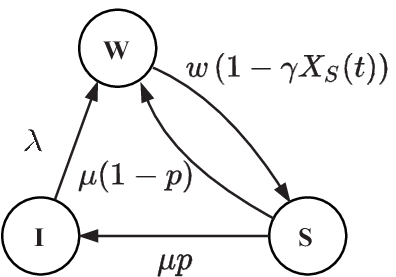}
\subcaption{Policy (W)}\label{fig:CTMC-policy-W}
\end{minipage}
\begin{minipage}[h]{0.325\linewidth}
\centering
        \includegraphics[scale=1]{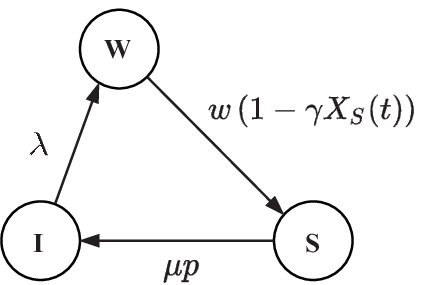}
\subcaption{Policy (S)}\label{fig:CTMC-policy-S}
\end{minipage}
\vspace{-0.1cm}
 \caption{\small{Illustration of the state transition diagram of the CTMC of each  device.}}
 \label{fig:ctmc}
\vspace{-0.35cm}
\end{figure}

In Figs.~\ref{fig:CTMC-policy-I} - \ref{fig:CTMC-policy-S}, we illustrate the state transition diagram of the CTMC of each IoT device under policies (I), (W), and (S), respectively. 
Let $\bs{\pi}\triangleq(\pi_d)_{d\in\mathcal{D}}$ be the stationary distribution of the CTMC of each device. 
 From Fig.~\ref{fig:ctmc}, we observe that the CTMCs are non-homogeneous with time-varying transition rates,
 and, thus, it is generally impossible to derive each corresponding  stationary distribution. 
 Later, we will derive explicit and accurate approximate expressions of these stationary distributions by using a mean-field approximation.

\subsection{Average AoI Metric}
We use the AoI as the performance metric to measure the freshness of the status information at the receiver for each IoT device. 
For each device, the instantaneous AoI at the associated receiver at time $t$ is defined as\cite{8469047}:
$\Delta(t) = t -U(t)$,
where 
 $U(t) $ is the timestamp of the most recently received status packet at time $t$.
Then, the average AoI of each device is defined as:
\begin{align}
\bar{\Delta} \triangleq \lim_{\tau\to\infty}\frac{1}{\tau}\int_{0}^{\tau}\Delta(t).
\end{align}

 In the following sections, we first analyze the average AoI of each device for the policies (I), (W), and (S) with schemes WP and WOP, under
 a given stationary distribution $\bs{\pi}$.
Then, we develop a mean-field approach to characterize the stationary distributions $\bs{\pi}$ under the three policies for the considered IoT system with a large number of devices.
These analytical results will allow us to better understand the effects of the system parameters on the AoI performance in the mean-field regime and gain design insights for practical ultra-dense IoT systems.

\section{Average AoI Analysis}\label{sec:aoi-analysis}
In this section, we derive the closed-form expression of the average AoI of each device for the policies defined in Section~\ref{sec:systemmodel}. 
Note that the standard graphical approach for calculating the average AoI may only be applicable for relatively simple systems, and, thus, could be very challenging for the considered complex system in which status packets can be dropped and the effects of transmission feedback need to be considered. 
To overcome the challenge, we adopt the SHS approach \cite{8469047} to evaluate the average AoI of each device in our system. First, we briefly introduce the main idea of the SHS approach for the average AoI analysis.

\subsection{Preliminaries on SHS}
For AoI analysis, the SHS approach models a system with hybrid state $(q(t),\bs{z}(t))$, where $q(t)\in\mathcal{Q}=\{0,\cdots,Q\}$ is a finite-state CTMC that describes the discrete state of the system and the age vector $\bs{z}(t)=[z_0(t)\cdots z_n(t)]\in\mathcal{R}^{1\times (n+1)}$ is a continuous process that describes the continuous-time evolution of age-related processes.

The CTMC $q(t)$ can be represented graphically as $(\mathcal{Q},\mathcal{L})$, where each state $q\in\mathcal{Q}$ is a node and each directed edge $l\triangleq(q_l,q_l')\in\mathcal{L}$ indicates a transition from state $q_l$ to state $q_l'$ with a transition rate $\lambda^{(l)}\delta_{q_l,q(t)}$. 
Here, the Kronecker delta function $\delta_{q_l,q(t)}$ ensures that the transition $l$ occurs  only when $q(t)$ equals to $q_l$.
Associated with each transition $l$, there is a transition reset mapping that leads to a discontinuous jump in the continuous process $\bs{z}(t)$.
Specifically, if a transition $l$ occurs, then the discrete state $q_l$ jumps to state $q_l'$ and the continuous state $\bs{z}$ is reset to $\bs{z}' =  \bs{z}\bs{A}_l$, where $\bs{A}_l\in\{0,1\}^{(n+1)\times (n+1)}$ is a binary transition reset map matrix.
Moreover, at each discrete state $q$, the continuous state $\bs{z}(t)$ evolves as a piecewise linear function through the differential equation 
$\dot{\bs{z}}(t) = \bs{b}_q$, where $\bs{b}_q\triangleq [b_{q,0}\cdots b_{q,n}]\in\{0,1\}^{1\times (n+1)}$ is a binary vector.
Here, $b_{q,j}= 1$  means that $z_j(t)$ increases at a unit rate in state $q$ and $b_{q,j}= 0$ indicates that $z_j(t)$ is irrelevant in state $q$ and does not need to be tracked. 

Note that, unlike an ordinary CTMC, the SHS Markov chain $(\mathcal{Q},\mathcal{L})$ may include self-transitions, during which, the discrete state $q(t)$ remains unchanged while a reset in the continuous state $\bs{z}(t)$ occurs. Moreover, for a given pair of discrete states $(q,q')$, there might be multiple transitions $l$ and $l'$ for which the transition reset map matrices $\bs{A}_l$ and $\bs{A}_{l'}$ are different.
 These two differences will be shown next while deriving the average AoI for our considered system.

For each state $q\in\mathcal{Q}$, let $\mathcal{L}_q'$ and $\mathcal{L}_q$ be, respectively, the set of incoming and outgoing transitions, i.e.,
 $\mathcal{L}_q'\triangleq\{l\in\mathcal{L}:q_l' = q\}$ and  $\mathcal{L}_q\triangleq\{l\in\mathcal{L}:q_l = q\}$.
 According to \cite[Theorem 4]{8469047}, if the CTMC $q(t)$ is ergodic with the stationary distribution  $\bs{\pi}\triangleq[\pi_0 \cdots\pi_Q]$, and there exists a non-negative vector $\bs{v} \triangleq [\bs{v}_0 \cdots\bs{v}_Q]$ with $\bs{v}_q\triangleq [v_{q0}\cdots v_{qM}]$ such that  
 \begin{align}
 \bs{v}_q \sum_{l\in\mathcal{L}_q} \lambda^{(l)} = \bs{b}_q \pi_q + \sum_{l\in\mathcal{L}_q'}\lambda^{(l)}\bs{v}_{q_l}\bs{A}_l, \forall q\in\mathcal{Q},\label{eqn:v_vector_yates}
 \end{align}
 then the average AoI is
 \begin{align}
 \bar{\Delta} =\sum_{q\in\mathcal{Q}} v_{q0}.\label{eqn:avg_aoi_yates}
 \end{align} 

\subsection{Characterization of Average AoI}
We now use the SHS approach to derive the closed-form expressions of the average AoI for our considered system under the stationary distribution $\bs{\pi}$ with policies (I), (W), and (S) for both the WP and WOP schemes. Such expressions have not been previously characterized in existing works \cite{bedewy2019optimizing,8901143,9007478,zhou2020meanfield} and they are useful for a better understanding of the AoI performance in ultra-dense IoT systems.

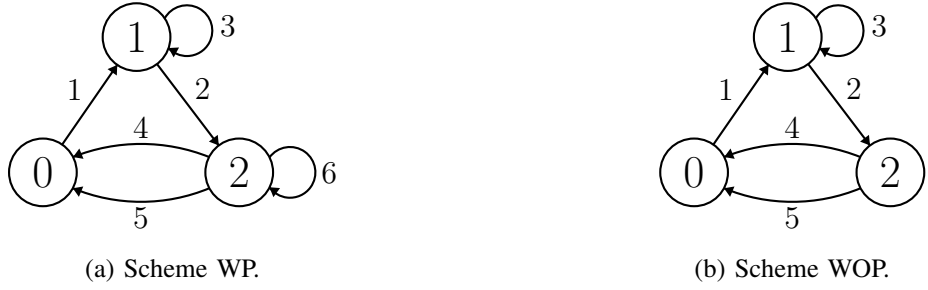
\begin{figure}[!t]
\centering
\begin{minipage}[h]{.49\linewidth}{}
\centering
\begin{tikzpicture}[scale=0.15]
\tikzstyle{every node}+=[inner sep=0pt]
\draw [thick,black] (22.5,-43.8) circle (3);
\draw (22.5,-43.8) node   {\Large $0$};
\draw [thick,black] (30.8,-31.8) circle (3);
\draw (30.8,-31.8) node {\Large $1$};
\draw [thick,black] (39.9,-43.8) circle (3);
\draw (39.9,-43.8) node {\Large $2$};
\draw [thick,black] (24.21,-41.33) -- (29.09,-34.27);
\fill [thick,black] (29.09,-34.27) -- (28.23,-34.64) -- (29.05,-35.21);
\draw (26.05,-36.44) node [left] {$1$};
\draw [thick,black] (32.61,-34.19) -- (38.09,-41.41);
\fill [thick,black] (38.09,-41.41) -- (38,-40.47) -- (37.21,-41.07);
\draw (35.92,-36.4) node [right] {$2$};
\draw [thick,black] (25.168,-42.437) arc (111.77128:68.22872:16.263);
\fill [thick,black] (25.17,-42.44) -- (26.1,-42.6) -- (25.73,-41.68);
\draw (31.2,-40.78) node [above] {$4$};
\draw [thick,black] (33.291,-30.149) arc (151.26898:-136.73102:2.25);
\draw (38.17,-30.81) node [right] {$3$};
\fill [thick,black] (33.63,-32.77) -- (34.09,-33.6) -- (34.57,-32.72);
\draw [thick,black] (42.58,-42.477) arc (144:-144:2.25);
\draw (47.15,-43.8) node [right] {$6$};
\fill [thick,black] (42.58,-45.12) -- (42.93,-46) -- (43.52,-45.19);
\draw [thick,black] (37.255,-45.205) arc (-67.46499:-112.53501:15.798);
\fill [thick,black] (25.15,-45.21) -- (25.69,-45.97) -- (26.08,-45.05);
\draw (31.2,-46.91) node [below] {$5$};
\end{tikzpicture}
 \subcaption{Scheme WP.}\label{fig:shs_ctmc_I_wp}
\end{minipage}
\begin{minipage}[h]{.49\linewidth}
\centering
\begin{tikzpicture}[scale=0.15]
\tikzstyle{every node}+=[inner sep=0pt]
\draw [thick,black] (22.5,-43.8) circle (3);
\draw (22.5,-43.8) node   {\Large $0$};
\draw [thick,black] (30.8,-31.8) circle (3);
\draw (30.8,-31.8) node {\Large $1$};
\draw [thick,black] (39.9,-43.8) circle (3);
\draw (39.9,-43.8) node {\Large $2$};
\draw [thick,black] (24.21,-41.33) -- (29.09,-34.27);
\fill [thick,black] (29.09,-34.27) -- (28.23,-34.64) -- (29.05,-35.21);
\draw (26.05,-36.44) node [left] {$1$};
\draw [thick,black] (32.61,-34.19) -- (38.09,-41.41);
\fill [thick,black] (38.09,-41.41) -- (38,-40.47) -- (37.21,-41.07);
\draw (35.92,-36.4) node [right] {$2$};
\draw [thick,black] (25.168,-42.437) arc (111.77128:68.22872:16.263);
\fill [thick,black] (25.17,-42.44) -- (26.1,-42.6) -- (25.73,-41.68);
\draw (31.2,-40.78) node [above] {$4$};
\draw [thick,black] (33.291,-30.149) arc (151.26898:-136.73102:2.25);
\draw (38.17,-30.81) node [right] {$3$};
\fill [thick,black] (33.63,-32.77) -- (34.09,-33.6) -- (34.57,-32.72);
\draw [thick,black] (37.255,-45.205) arc (-67.46499:-112.53501:15.798);
\fill [thick,black] (25.15,-45.21) -- (25.69,-45.97) -- (26.08,-45.05);
\draw (31.2,-46.91) node [below] {$5$};
\end{tikzpicture}
 \subcaption{Scheme WOP.}\label{fig:shs_ctmc_I_wop}
\end{minipage}
\vspace{-0.2cm}
 \caption{\small{Illustration of the SHS Markov chain under policy (I).}}
 \label{fig:shs-ctmc-I}
\vspace{-0.35cm}
\end{figure}

We begin with policy (I) under scheme WP. For a given stationary distribution $\bs{\pi}$, the transition rate from state \textit{W} to state \textit{I} is $w(1-\gamma \pi_{\textit{S}})$. Let $k\triangleq w(1-\gamma \pi_{\textit{S}})$ be the effective waiting rate.
To model the state of each device in our system under policy (I) with scheme WP by using the SHS approach, the discrete state is $q(t)\in\mathcal{Q}=\{0,1,2\}$ where $0$, $1$, and $2$  indicate states \textit{I}, \textit{W}, and \textit{S}, respectively,
and the continuous state is $\bs{z}(t) = [z_0(t)~z_1(t)]$, where $z_0(t)$ is the current AoI $\Delta(t)$ at the receiver and  $z_1(t)$ is the age of the packet at the device, either in waiting or in service.
The SHS Markov chain of $q(t)$ is illustrated in Fig.~\ref{fig:shs_ctmc_I_wp} and  the corresponding transitions are summarized in Table~\ref{table:shs_ctms_I_wp} and explained next:
    \begin{itemize}
        \item $l=1$: A status packet arrives at the device in state \textit{I}. With this arrival, $z_0'=z_0$ is unchanged because it does yield an AoI reduction at the receiver. However, $z_1'=0$ as the newly arriving status packet is fresh and its age is zero. 
        \item $l=2$: The device finds an idle channel and finishes the waiting period. This does not change the AoI at the receiver nor the age of the device's current packet. Thus, $z_0'=z_0$ and $z_1'=z_1$.
        \item $l=3$: A status packet arrives at the device in state \textit{W}. This does not change the AoI at the receiver, and, thus $z_0'=z_0$. However, this new status packet with age zero will replace the packet currently at the device. Thus, $z_1'=0$.
        \item $l=4$: The device completes its service and successfully delivers a packet to the receiver. With this transition, the AoI at the receiver is reset to the age of the packet at the device that is delivered, i.e., $z_0'=z_1$. Since there is no packet at the device,  $z_1'$ is set to zero.
        \item $l=5$: The device completes its service and fails to deliver a status packet to the receiver. Thus,  the AoI at the receiver and the age of the packet at the device remain unchanged, i.e., $z_0'=z_0$ and $z_1'=z_1$.
        \item $l=6$: The packet in service at the device is preempted by a newly arriving status packet under  scheme WP. The AoI at the receiver remains unchanged while the age of the packet at the device is reset to zero, since the new packet is fresh, i.e., $z_0'=z_0$ and $z_1'=0$.
    \end{itemize}

Clearly, the SHS Markov chain in Fig.~\ref{fig:shs_ctmc_I_wp} is different from an ordinary CTMC in Fig.~\ref{fig:CTMC-policy-I}.
In particular, the SHS Markov chain in Fig.~\ref{fig:shs_ctmc_I_wp} includes self-transitions (at states $1$ and $2$) during which the discrete state $q(t)$ remains unchanged while a reset in the continuous state $\bs{z}(t)$ occurs. Moreover, for a given pair of discrete states $0$ and $2$, there are two transitions $l=4$ and $l'=5$ for which the transition reset map matrices $\bs{A}_4$ and $\bs{A}_{5}$ are different.

\begin{table}[!t]
\centering
\caption{\small{Transitions for the SHS Markov chain under policy (I) with scheme WP in Fig.~\ref{fig:shs_ctmc_I_wp}.}}\label{table:shs_ctms_I_wp}
\begin{tabular}{cccccc}
\hline
$l$ & $q_l\to q_l'$ & $\lambda^{l}$  & $\bs{z}\bs{A}_l$ & $\bs{A}_l$  & $\bs{v}_{q_l} \bs{A}_l$ \\
\hline
$1$ & $0 \to 1$     & $\lambda$   & $\begin{bmatrix} z_0 & 0 \end{bmatrix}$ & $\begin{bmatrix} 1 & 0 \\ 0 & 0 \end{bmatrix}$ & $\begin{bmatrix} \bar{v}_{00} & 0 \end{bmatrix}$    \\         
$2$ & $1 \to 2$     & $k$   & $\begin{bmatrix} z_0 & z_1 \end{bmatrix}$ & $\begin{bmatrix} 1 & 0 \\ 0 & 1 \end{bmatrix}$ & $\begin{bmatrix} \bar{v}_{10} & \bar{v}_{11} \end{bmatrix}$    \\  
$3$ & $1 \to 1$     & $\lambda$     & $\begin{bmatrix} z_0 & 0 \end{bmatrix}$ & $\begin{bmatrix} 1 & 0 \\ 0 & 0 \end{bmatrix}$ & $\begin{bmatrix} \bar{v}_{10} & 0 \end{bmatrix}$    \\         
$4$ & $2 \to 0$     & $\mu p$ & $\begin{bmatrix} z_1 & 0 \end{bmatrix}$ & $\begin{bmatrix} 0 & 0 \\ 1 & 0 \end{bmatrix}$ & $\begin{bmatrix} \bar{v}_{21} & 0 \end{bmatrix}$    \\         
$5$ & $2 \to 0$     & $\mu (1-p)$   & $\begin{bmatrix} z_0 & z_1 \end{bmatrix}$ & $\begin{bmatrix} 1 & 0 \\ 0 & 1 \end{bmatrix}$ & $\begin{bmatrix} \bar{v}_{20} & \bar{v}_{21} \end{bmatrix}$    \\         
$6$ & $2 \to 2$     & $\lambda$  & $\begin{bmatrix} z_0 & 0 \end{bmatrix}$ & $\begin{bmatrix} 1 & 0 \\ 0 & 0 \end{bmatrix}$ & $\begin{bmatrix} \bar{v}_{20} & 0 \end{bmatrix}$     \\
\hline 
\end{tabular}
\vspace{-0.6cm}
\end{table}

The evolution of $\bs{z}(t)$ depends on the discrete state $q(t)$. In particular, when $q(t)=q$,  we have
\begin{equation}\label{eqn:bq}
\dot{\bs{z}}(t) = \bs{b}_q =  \left\{
 \left.\begin{aligned}
        &[1~0],\quad\text{if~} q=0,\\
        &[1~1],\quad\text{if~} q= 1,2.
       \end{aligned}
 \right.\right.
\end{equation}
\eqref{eqn:bq} holds because the AoI at the receiver $\Delta(t)=z_0(t)$ always increases at a unit rate with time $t$ in all discrete states, and the age of the packet at the device $z_1(t)$ increases at a unit rate in state $q=1,2$ in which there is a packet at the device.

From Fig.~\ref{fig:shs_ctmc_I_wp} and Table~\ref{table:shs_ctms_I_wp}, the stationary distribution $\bs{\pi}=[\pi_0~ \pi_1 ~\pi_2]$ of $q(t)$ satisfies:
\begin{subequations}
\begin{align}
&\lambda \pi_0 = k \pi_1 = \mu \pi_2,\label{eqn:pi_I_1}\\
&\pi_0 + \pi_1 + \pi_2 = 1.\label{eqn:pi_I_2}
\end{align}
\end{subequations}
Then, we have:
\begin{align}\label{eqn:pi_i}
\bs{\pi}=[\pi_0~ \pi_1 ~\pi_2] = \frac{1}{k\mu + \lambda\mu +k \lambda}\left[ k\mu ~ \lambda\mu ~ k \lambda\right].
\end{align}
Next, we calculate $\bs{v} = [\bs{v}_0~\bs{v}_1~\bs{v}_2] =[v_{00}~v_{01}~v_{10}~v_{11}~v_{20}~v_{21} ]$ in \eqref{eqn:v_vector_yates}. 
By Table~\ref{table:shs_ctms_I_wp} and \eqref{eqn:bq}, we have\begin{subequations}\label{eqn:vq_I_WP}
\begin{align}
&\lambda [v_{00}~v_{01}] = \pi_0[1~0] + \mu p [v_{21}~0] + \mu(1-p) [v_{20}~v_{21}],\\
&(k+\lambda) [v_{10}~v_{11}] = \pi_1[1~1]  + \lambda [v_{00}~0] + \lambda[v_{10}~0],\\
&(\mu+\lambda) [v_{20}~v_{21}] = \pi_2[1~1]  + k [v_{10}~v_{11}] + \lambda[v_{20}~0].
\end{align}
\end{subequations}
According to \eqref{eqn:avg_aoi_yates}, after complex calculations of \eqref{eqn:vq_I_WP} (given in Appendix~A), we can obtain the closed-form expression of the average AoI under policy (I) with scheme WP, given by
\begin{align}
\bar{\Delta}_{\textrm{I-WP}}=v_{00}+v_{01} + v_{11}.
\end{align}

For policy (I) with scheme WOP, the analysis is similar. In particular, the corresponding SHS Markov chain is illustrated in Fig.~\ref{fig:shs_ctmc_I_wop}. We can see that, compared with the SHS Markov chain for policy (I) with scheme WP in Fig.~\ref{fig:shs_ctmc_I_wp}, there is no transition $l=6$ for the SHS Markov chain in Fig.~\ref{fig:shs_ctmc_I_wop}, since preemption in service is not allowed. Here, the  transitions $l=1,2,\cdots,5$ are the same as those in Table~\ref{table:shs_ctms_I_wp}. We can further see that the stationary distribution $\bs{\pi}=[\pi_0~ \pi_1 ~\pi_2]$ is the same to the one in \eqref{eqn:pi_i}. Then, according to \eqref{eqn:v_vector_yates} and \eqref{eqn:avg_aoi_yates}, we can obtain $\bar{\Delta}_{\textrm{I-WOP}}=v_{00}+v_{01} + v_{11}$,
where $v_{00}$, $v_{01}$, and $v_{11}$ satisfy the following system of linear equations:
\begin{subequations}\label{eqn:vq_I_WOP}
\begin{align}
&\lambda [v_{00}~v_{01}] = \pi_0[1~0] + \mu p [v_{21}~0] + \mu(1-p) [v_{20}~v_{21}],\\
&(k+\lambda) [v_{10}~v_{11}] = \pi_1[1~1]  + \lambda [v_{00}~0] + \lambda[v_{10}~0],\\
&(\mu+\lambda) [v_{20}~v_{21}] = \pi_2[1~1]  + k [v_{10}~v_{11}].
\end{align}
\end{subequations}


For policies (W) and (S), based on the SHS approach, proceeding in a similar way to that used for policy (I), we can also derive the closed-form expressions of the average AoI.
In the following theorem, we summarize the closed-form expressions of the average AoI of each device for the three policies.
\begin{theorem} \label{theorem:aoi}
Under the stationary distribution $\bs{\pi}$,  for each device, the average AoI under  policies (I), (W), and (S) with schemes WP and WOP are given as follows.
\begin{itemize}
\item Under policy (I), we have:
\begin{align}
&\bar{\Delta}_{\textrm{I-WP}} = \left(\frac{1}{\lambda} + \frac{1}{k} + \frac{1}{\mu}\right)\frac{1}{p} + \frac{\lambda + k +\mu}{(\lambda+\mu)(\lambda + k) } - \frac{\lambda + k + \mu}{\lambda k + k\mu + \lambda\mu},\label{eqn:avg_aoi_wp-I}\\
&\bar{\Delta}_{\textrm{I-WOP}} = \left(\frac{1}{\lambda} + \frac{1}{k} + \frac{1}{\mu}\right)\frac{1}{p} + \frac{1}{\mu} +  \frac{1}{\lambda + k } - \frac{\lambda + k + \mu}{\lambda k + k\mu + \lambda\mu}.\label{eqn:avg_aoi_wop-I}
\end{align}

\item Under policy (W), we have:
\begin{align}
&\bar{\Delta}_{\textrm{W-WP}} = \left(\frac{p}{\lambda} + \frac{1}{k} + \frac{1}{\mu}\right)\frac{1}{p} + \frac{\lambda + k + \mu}{(\lambda+\mu)(k+\lambda)-k\mu (1-p)}- \frac{\lambda + k + \mu}{\lambda k + \lambda\mu + k\mu p}  ,\label{eqn:avg_aoi_wp-W}\\
&\bar{\Delta}_{\textrm{W-WOP}} = \left(\frac{p}{\lambda} + \frac{1}{k} + \frac{1}{\mu}\right)\frac{1}{p} + \frac{\lambda + k + \mu}{\mu(kp+\lambda)}  - \frac{\lambda + k + \mu}{\lambda k + \lambda\mu + k\mu p}.\label{eqn:avg_aoi_wop-W}
\end{align}
\item Under policy (S), we have:
\begin{align}
&\bar{\Delta}_{\textrm{S-WP}} = \frac{1}{\lambda} + \frac{1}{k} + \frac{1}{\mu p} + \frac{\mu p + k + \lambda}{(\lambda + \mu p)(\lambda + k)} - \frac{\lambda + k + \mu p}{\lambda k + (k + \lambda)\mu p},\label{eqn:avg_aoi_wp-S}\\
&\bar{\Delta}_{\textrm{S-WOP}} = \frac{1}{\lambda} + \frac{1}{k} + \frac{2}{\mu p} + \frac{1}{\lambda + k } - \frac{\lambda + k + \mu p}{\lambda k + (k + \lambda)\mu p}.\label{eqn:avg_aoi_wop-S}
\end{align}
\end{itemize}
\end{theorem}

\begin{IEEEproof}
  See Appendix~A.
\end{IEEEproof}

From Theorem~\ref{theorem:aoi}, we observe that, for the case with error-free channels, i.e., $p=1$, we have: $\bar{\Delta}_{\textrm{I-WP}} = \bar{\Delta}_{\textrm{W-WP}} = \bar{\Delta}_{\textrm{S-WP}}$ and $\bar{\Delta}_{\textrm{I-WOP}} = \bar{\Delta}_{\textrm{W-WOP}} = \bar{\Delta}_{\textrm{S-WOP}}$, which are consistent with the expressions of the average AoI in \cite[Theorem 1]{zhou2020meanfield}.
Moreover, for each of of the three policies, we show that scheme WP always achieves a smaller average AoI compared with scheme WOP, as summarized in the following lemma.
\begin{lemma}
Under a given stationary distribution $\bs{\pi}$, we have $\bar{\Delta}_{\textrm{I-WP}}< \bar{\Delta}_{\textrm{I-WOP}}$, $\bar{\Delta}_{\textrm{W-WP}}< \bar{\Delta}_{\textrm{W-WOP}}$, and $\bar{\Delta}_{\textrm{S-WP}}< \bar{\Delta}_{\textrm{S-WOP}}$.
\end{lemma} 

\begin{IEEEproof}
Based on Theorem~\ref{theorem:aoi}, under a given $\bs{\pi}$, we have:
\begin{align*}
&\bar{\Delta}_{\textrm{I-WOP}} - \bar{\Delta}_{\textrm{I-WP}} = \frac{1}{\mu} +  \frac{1}{\lambda + k }  - \frac{\lambda + k +\mu}{(\lambda+\mu)(\lambda + k) } = \frac{\lambda^2 + 2\lambda\mu}{\mu(\lambda+\mu)(\lambda + k)}>0,\\
&\bar{\Delta}_{\textrm{W-WOP}} - \bar{\Delta}_{\textrm{W-WP}} = \frac{\lambda + k + \mu}{\mu(kp+\lambda)}   - \frac{\lambda + k + \mu}{(\lambda+\mu)(k+\lambda)-k\mu (1-p)} \nonumber\\
&\hspace{30mm}= \frac{\lambda(\lambda + k + \mu)(k+\lambda)}{\mu(kp+\lambda)(k(\lambda+\mu p) + \mu(\lambda +\mu))}>0,\\
&\bar{\Delta}_{\textrm{S-WOP}} - \bar{\Delta}_{\textrm{S-WP}} = \frac{1}{\mu p} + \frac{1}{\lambda + k }  - \frac{\mu p + k + \lambda}{(\lambda + \mu p)(\lambda + k)} = \frac{\lambda^2 + 2\lambda\mu p}{\mu p (\lambda+\mu p)(\lambda + k)}>0,
\end{align*}
which completes the proof.
\end{IEEEproof}

Note that, we cannot directly compare the average AoI among the three policies, by using the closed-form expressions under a given stationary distribution in Theorem~\ref{theorem:aoi}. The reason is that under the same system parameters, the stationary distributions of the three policies would be different, as can be seen from the CTMCs in Fig.~\ref{fig:ctmc}. Later, in the simulations, we will compare the average AoI for these three policies. 
Theorem~\ref{theorem:aoi} provides rigorous analytical characterizations of the average AoI under a given stationary distribution $\bs{\pi}$.
Next, we analyze $\bs{\pi}$ under the three policies for the considered IoT monitoring system under an ultra-dense regime with a large number of devices by a mean-field approximation \cite{10.1145/3084454,10.1145/2964791.2901463,10.1145/2825236.2825241}.

\section{Mean-Field Analysis for Dense IoT}\label{sec:mean-field}
Next, we develop a mean-field framework to characterize the stationary distributions $\bs{\pi}$ under policies (I), (W) and (S), for the considered system in the mean-field regime, where the number of devices $N$ goes to infinity and $\gamma$ is a constant.
Then, we can replace the complex non-homogeneous CTMC $D_n(t)$ by a much simpler deterministic dynamic system, and then  obtain an explicit approximate expression of the stationary distribution $\bs{\pi}$. 

Note that, by comparing the CTMC in Fig.~\ref{fig:CTMC-policy-I} with the CTMC studied in \cite[Fig. 1(b)]{zhou2020meanfield}, we can see that the corresponding mean-field analysis is analogous to  \cite[Section IV-A]{zhou2020meanfield}.
Thus, we omit the analysis of policy (I) due to space limitations.

For policy (W), for a finite $N$, based on Fig.~\ref{fig:CTMC-policy-W}, we obtain the following transitions of $\bs{X}(t)$: 
\begin{equation}\label{eqn:transition_X}
 \left\{\left.\begin{aligned}
        &\bs{X} \mapsto \bs{X} + \frac{1}{N}(-1,1,0)~\text{at~rate}~N\lambda X_{\textit{I}},\\
        &\bs{X} \mapsto \bs{X} + \frac{1}{N}(0,-1,1)~\text{at~rate}~Nw(1-\gamma X_{\textit{S}}) X_{\textit{W}} -N\mu(1-p)X_{\textit{S}},\\
        &\bs{X} \mapsto \bs{X} + \frac{1}{N}(1,0,-1)~\text{at~rate}~N\mu p X_{\textit{S}}.
       \end{aligned}
 \right.\right.
\end{equation}

From the form of the transitions in \eqref{eqn:transition_X}, we know that $\bs{X}(t) $ belongs to the class of density-dependent population processes \cite{10.1145/3084454} and, hence, we  can characterize the corresponding mean-field model ($N\to\infty$) by an ordinary differential equation (ODE)
$\dot{\bs{x}} = f(\bs{x})$. Here, $\bs{x} \triangleq (x_{\textit{I}},x_{\textit{W}},x_{\textit{S}})$ and  $f(\bs{x}) \triangleq \lim_{dt\to 0} \left(\mathbb{E}[\bs{X}(t+dt)-\bs{X}(t)\mid \bs{X}(t)=\bs{x}]\right)/dt$ is the drift. 
By \eqref{eqn:transition_X}, we have
\begin{equation}\label{eqn:mean-field}
 \left\{
 \left.\begin{aligned}
        &\dot{x}_{\textit{I}} = -\lambda x_{\textit{I}} + \mu p x_{\textit{S}},\\
        &\dot{x}_{\textit{W}} =  \lambda x_{\textit{I}} -  w(1-\gamma x_{\textit{S}}) x_{\textit{W}} + \mu (1-p) x_{\textit{S}},\\
        &\dot{x}_{\textit{S}} = w(1-\gamma x_{\textit{S}}) x_{\textit{W}} - \mu x_{\textit{S}}.
       \end{aligned}
 \right.\right.
\end{equation}
Let $\bs{x}^*$ be an equilibrium point of the ODE in \eqref{eqn:mean-field}. We now show the mean-field approximation in \eqref{eqn:mean-field} is accurate for the considered system under policy (W) in the following theorem. 

\begin{theorem}\label{theorem:mean-field}
Under policy (W), the equilibrium point $\bs{x}^*$ of the mean-field model in \eqref{eqn:mean-field} is unique, and satisfies:
\begin{subequations}\label{eqn:mean-field-solution}
\begin{align}
&x_{\textit{I}}^* = \frac{\mu p}{\lambda}x_{\textit{S}}^*,~x_{\textit{W}}^* = \frac{\mu x_{\textit{S}}^*}{w(1-\gamma x_{\textit{S}}^*)},\\
&x_{\textit{S}}^* = \dfrac{w(\lambda + \mu p + \lambda \gamma) + \lambda\mu  - \sqrt{(w(\lambda + \mu p + \lambda \gamma) + \lambda\mu)^2 -4\lambda (\lambda + \mu p)\gamma w^2}}{2w\gamma(\lambda+\mu p)}.\label{eqn:xs_theorem2}
\end{align}
\end{subequations}

As $N$ goes to infinity, the stationary distribution $\bs{\pi}$ and the average AoI under schemes WP and WOP, $\bar{\Delta}_{\textrm{W-WP}}(\bs{\pi})$ and $\bar{\Delta}_{\textrm{W-WOP}}(\bs{\pi})$, converge, respectively, to $\bs{x}^*$,  $\bar{\Delta}_{\textrm{W-WP}}(\bs{x}^*)$, and  $\bar{\Delta}_{\textrm{W-WOP}}(\bs{x}^*)$, with the rates of convergence:
\begin{subequations}\label{eqn:rates_of_convergence}
\begin{align}
&\left|\mathbb{E}[\bs{\pi}] - \bs{x}^*\right| = O(\frac{1}{N}),\\
&\left|\mathbb{E}[\bar{\Delta}_{\textrm{W-WP}}(\bs{\pi})] - \bar{\Delta}_{\textrm{W-WP}}(\bs{x}^*)\right| = O(\frac{1}{N}),\\
&\left|\mathbb{E}[\bar{\Delta}_{\textrm{W-WOP}}(\bs{\pi})] - \bar{\Delta}_{\textrm{W-WOP}}(\bs{x}^*)\right|= O(\frac{1}{N}),
\end{align} 
\end{subequations}
\noindent where the expectation is taken over the stationary distribution $\bs{\pi}$.
\end{theorem}
\begin{IEEEproof}
See Appendix~B.
\end{IEEEproof}

For policy (S), by comparing the associated CTMC  in \ref{fig:CTMC-policy-S} with the CTMC in Fig.~\ref{fig:CTMC-policy-I} under policy (I), we can see that the mean-field analysis can be done using \cite[Section IV-A]{zhou2020meanfield}, yielding the following theorem:

\begin{theorem}\label{theorem:mean-field-S}
Under policy (S), as the number of the devices $N$ goes to infinity and $\gamma = N/M$ remains a constant, the stationary distribution the stationary distribution $\bs{\pi}$ and the average AoI under schemes WP and WOP, $\bar{\Delta}_{\textrm{S-WP}}(\bs{\pi})$ and $\bar{\Delta}_{\textrm{S-WOP}}(\bs{\pi})$, converge, respectively, to $\bs{x}^*$,  $\bar{\Delta}_{\textrm{S-WP}}(\bs{x}^*)$, and  $\bar{\Delta}_{\textrm{S-WOP}}(\bs{x}^*)$, with the rates of convergence:
\begin{subequations}\label{eqn:rates_of_convergence-S}
\begin{align}
&\left|\mathbb{E}[\bs{\pi}] - \bs{x}^*\right| = O(\frac{1}{N}),\\
&\left|\mathbb{E}[\bar{\Delta}_{\textrm{S-WP}}(\bs{\pi})] - \bar{\Delta}_{\textrm{S-WP}}(\bs{x}^*)\right| = O(\frac{1}{N}),\\
&\left|\mathbb{E}[\bar{\Delta}_{\textrm{S-WOP}}(\bs{\pi})] - \bar{\Delta}_{\textrm{S-WOP}}(\bs{x}^*)\right|= O(\frac{1}{N}),
\end{align} 
\end{subequations}
where the expectation is taken over the stationary distribution $\bs{\pi}$ and $\bs{x}^*=(x_{\textit{I}}^*,x_{\textit{W}}^*,x_{\textit{S}}^*)$ satisfies:\begin{subequations}\label{eqn:mean-field-solution-S}
\begin{align}
&x_{\textit{I}}^* = \frac{\mu p}{\lambda}x_{\textit{S}}^*,~x_{\textit{W}}^* = \frac{\mu p x_{\textit{S}}^*}{w(1-\gamma x_{\textit{S}}^*)},\\
&x_{\textit{S}}^* = \dfrac{w(\lambda + \mu p + \lambda \gamma) + \lambda\mu p - \sqrt{(w(\lambda + \mu p + \lambda \gamma) + \lambda\mu p)^2 -4\lambda (\lambda + \mu p)\gamma w^2}}{2w\gamma(\lambda+\mu p)}.\label{eqn:xs_theorem3}
\end{align}
\end{subequations}
\end{theorem}

Next, we analyze the influence of the system parameters, i.e., $\lambda$, $\mu$, $w$, $p$, and $\gamma$, on the equilibrium point $\bs{x}^*$ under each of the three policies in the mean-field limit.
\begin{lemma}\label{lemma:properties_x}
Under policies (I), (W), and (S), the mean-field equilibrium point $\bs{x}^*$ has the following properties with respect to the system parameters:
\begin{itemize}
\item As the arrival rate $\lambda$ increases, under the three policies, $x_{\textit{I}}^*$ decreases, and  both $x_{\textit{W}}^*$ and $x_{\textit{S}}^*$ increase;
\item As the service rate $\mu$ increases, under the three policies, $x_{\textit{I}}^*$ increases and $x_{\textit{S}}^*$ decreases;
\item As the waiting rate $w$ increases, under the three policies, $x_{\textit{I}}^*$ increases, and  both $x_{\textit{W}}^*$ and $x_{\textit{S}}^*$ decrease;
\item As the successful transmission probability $p$ increases, under policy (I), $\bs{x}^*$ remains unchanged; under policy (W), $x_{\textit{I}}^*$ increases, and  both $x_{\textit{W}}^*$ and $x_{\textit{S}}^*$ decrease; and under policy (S),  $x_{\textit{I}}^*$ increases and $x_{\textit{S}}^*$ decreases;
\item As the ratio $\gamma$ increases,  under the three policies, $x_{\textit{I}}^*$ decreases, and  both $x_{\textit{W}}^*$ and $x_{\textit{S}}^*$ increase.
\end{itemize}
\end{lemma}
\begin{IEEEproof}
See Appendix~C. 
\end{IEEEproof}

From Lemma~\ref{lemma:properties_x}, we can see that, when the channel conditions are better (i.e., a larger $\mu$ or a larger $p$) or there are more communication resources (i.e., a smaller $\gamma$), then there will fewer devices in the service state. Meanwhile, when  the packet traffic is heavier (i.e., a larger $\lambda$) or the devices are more aggressive in accessing the channels (i.e., a large $w$), then there will more devices in the service state.  
Here, as seen in Lemma~\ref{lemma:properties_x}, $x_{\textit{W}}^*$ may not possess similar monotonicity properties with respect to $\mu$ under the three policies and with respect to $p$ under policy (S).
The properties obtained in Lemma~\ref{lemma:properties_x} will be exploited to investigate the impacts of the system parameters on the average AoI  in Theorem~\ref{theorem:aoi} in the mean-field limit.\footnote{These properties in Lemma~\ref{lemma:properties_x} can be used for further analytical investigations on other network performance metrics, for example, the average throughput and the average energy cost.} 

\begin{theorem}\label{theorem:properties-of-aoi}
Under both policies (I) and (S) with schemes WP and WOP, the average AoI in the mean-field limit, i.e., $\bar{\Delta}_{\textrm{I-WP}}(\bs{x}^*)$, $\bar{\Delta}_{\textrm{I-WOP}}(\bs{x}^*)$, $\bar{\Delta}_{\textrm{S-WP}}(\bs{x}^*)$, and $\bar{\Delta}_{\textrm{S-WOP}}(\bs{x}^*)$, decrease, with the increase of $\mu$, $w$, and $p$, and the decrease of $\gamma$.
\end{theorem}
\begin{IEEEproof}
See Appendix~D. 
\end{IEEEproof}

Theorem~\ref{theorem:properties-of-aoi} indicates that, under policies (I) and (S) with schemes WP and WOP, the considered IoT system with higher service rates, higher waiting rates, more communication resources, and better channel conditions, can achieve a smaller average AoI. 
Moreover, the associated average AoI does not necessarily decrease with the arrival rate, as will be shown in the simulations.
Here, for policy (W), due to its more complex expressions of the average AoI in \eqref{eqn:avg_aoi_wp-W} and \eqref{eqn:avg_aoi_wop-W}, similar properties as in Theorem~\ref{theorem:properties-of-aoi} cannot be shown theoretically and will be illustrated via simulations.

\section{Simulation Results and Analysis}\label{sec:simulations}
We now present numerical results to investigate  the average AoI of the three policies under a given $\bs{\pi}$ in Theorem~\ref{theorem:aoi}, the accuracy of the mean-field approximation for policy (W) in Theorem~\ref{theorem:mean-field}, and the average AoI of the three policies in the mean-field limit $\bs{x}^*$.
\begin{figure}[!t]
\begin{centering}
\includegraphics[scale=.45]{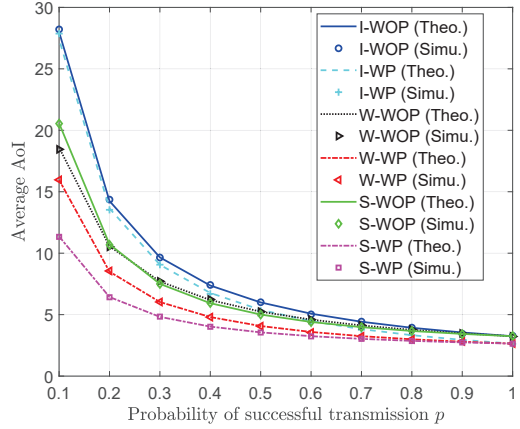}
\vspace{-0.1cm}
 \caption{\small{Average AoI versus the successful transmission probability $p$ under a given stationary distribution. $\lambda=0.9,\mu=1$ and $k=2$.}\label{fig:AoI_verify}}
\end{centering}
\vspace{-0.35cm}
\end{figure}

In Fig.~\ref{fig:AoI_verify}, we evaluate the analytical results in Theorem~\ref{theorem:aoi} and the simulations results of the average AoI under the three policies with schemes WP and WOP. The simulations results are obtained by averaging over 50,000 status packet arrivals. 
We can see that the simulation results agree very well with the analytical results thus corroborating the theoretical results characterized in Theorem~\ref{theorem:aoi}.  
We also observe that, for a fixed $k$, the average AoI of the three policies with and without preemption decrease with $p$.  This  corroborates the intuition that better channel conditions will achieve a smaller average AoI.

\begin{figure}[!t]
\centering
\begin{minipage}[h]{.32\linewidth}
\centering
       \includegraphics[scale=0.35]{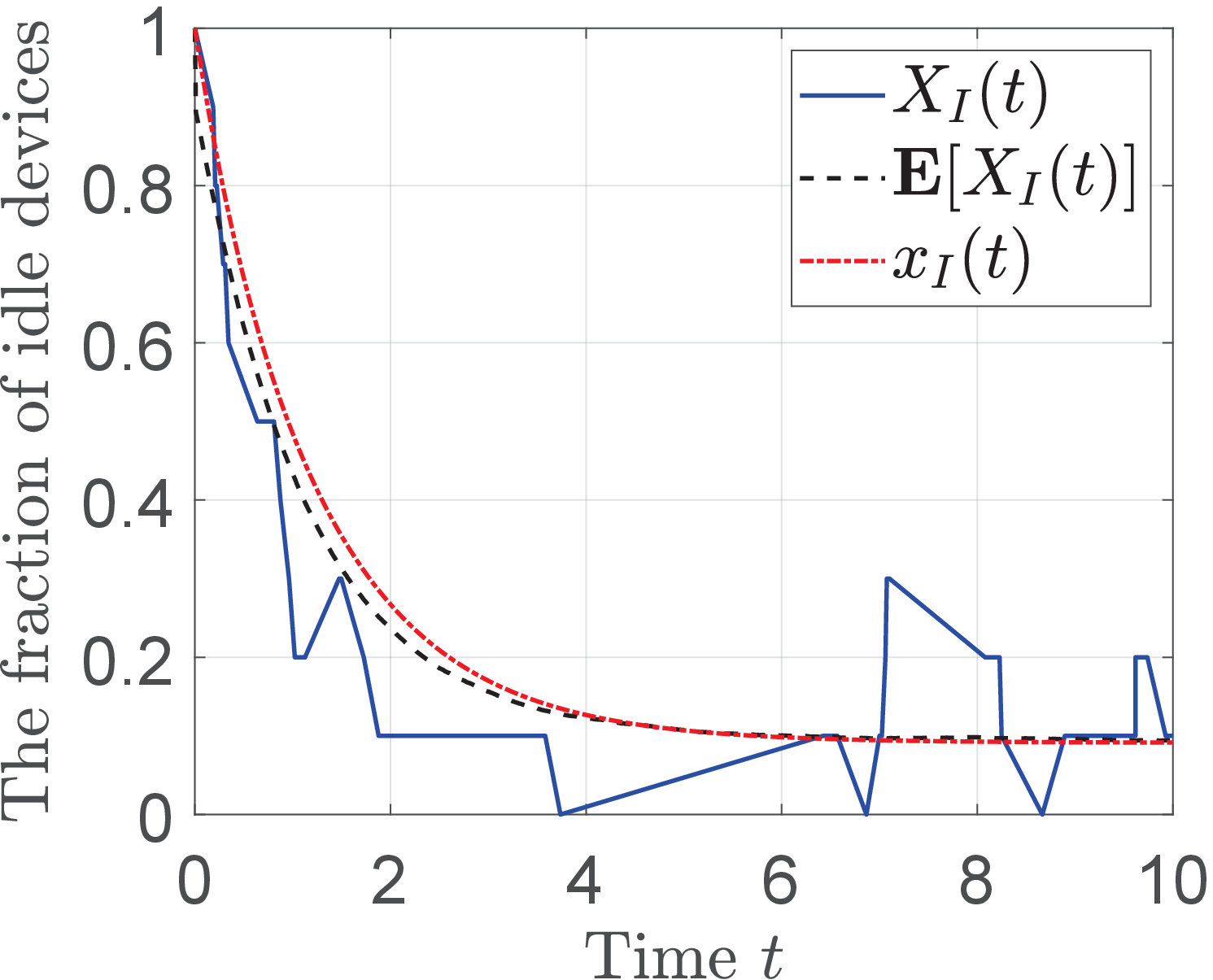}
\subcaption{$N=10$}\label{fig:N10}
\end{minipage}
\begin{minipage}[h]{.32\linewidth}
\centering
        \includegraphics[scale=0.35]{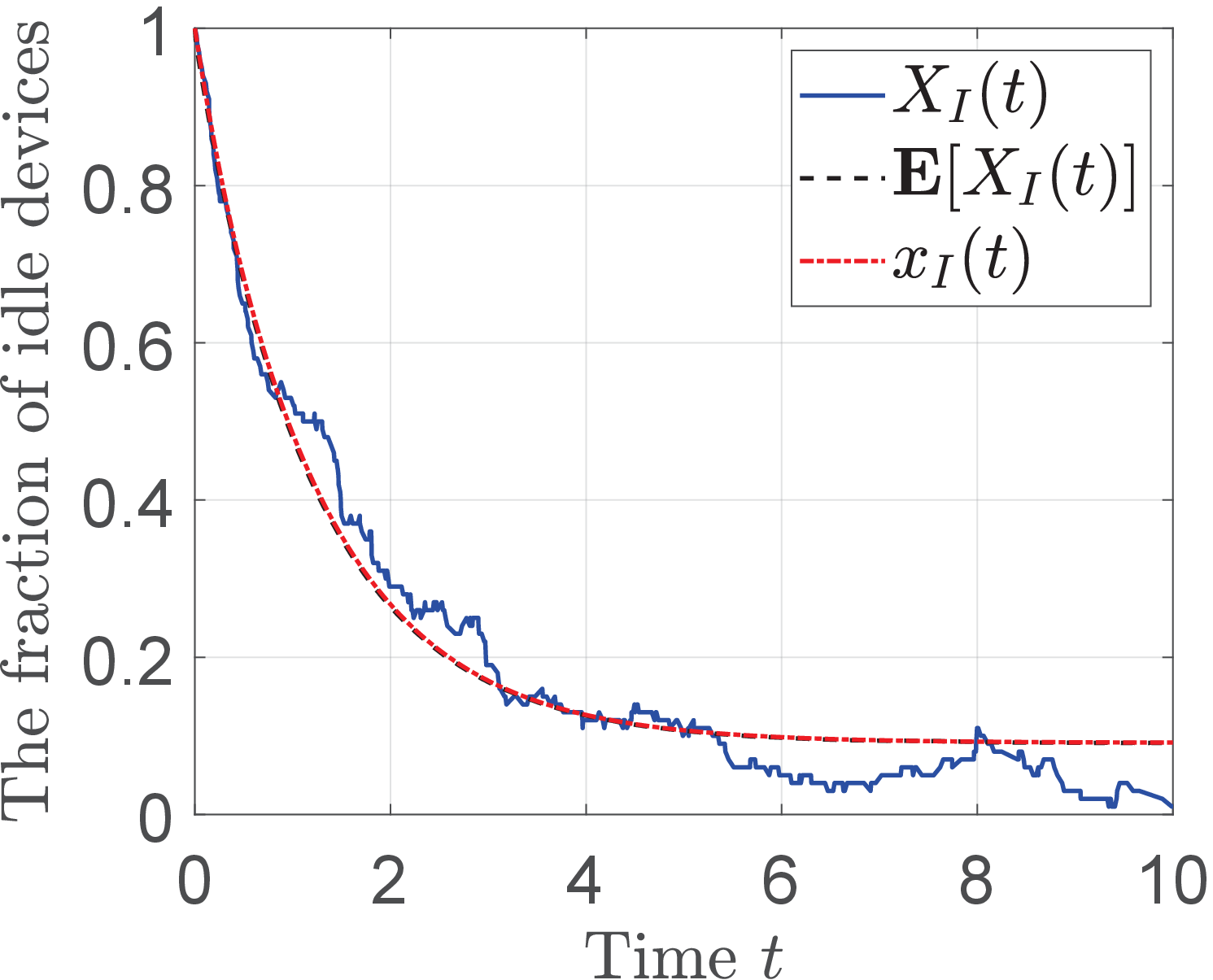}
\subcaption{ $N=100$}\label{fig:N100}
\end{minipage}
\begin{minipage}[h]{0.32\linewidth}
\centering
        \includegraphics[scale=0.35]{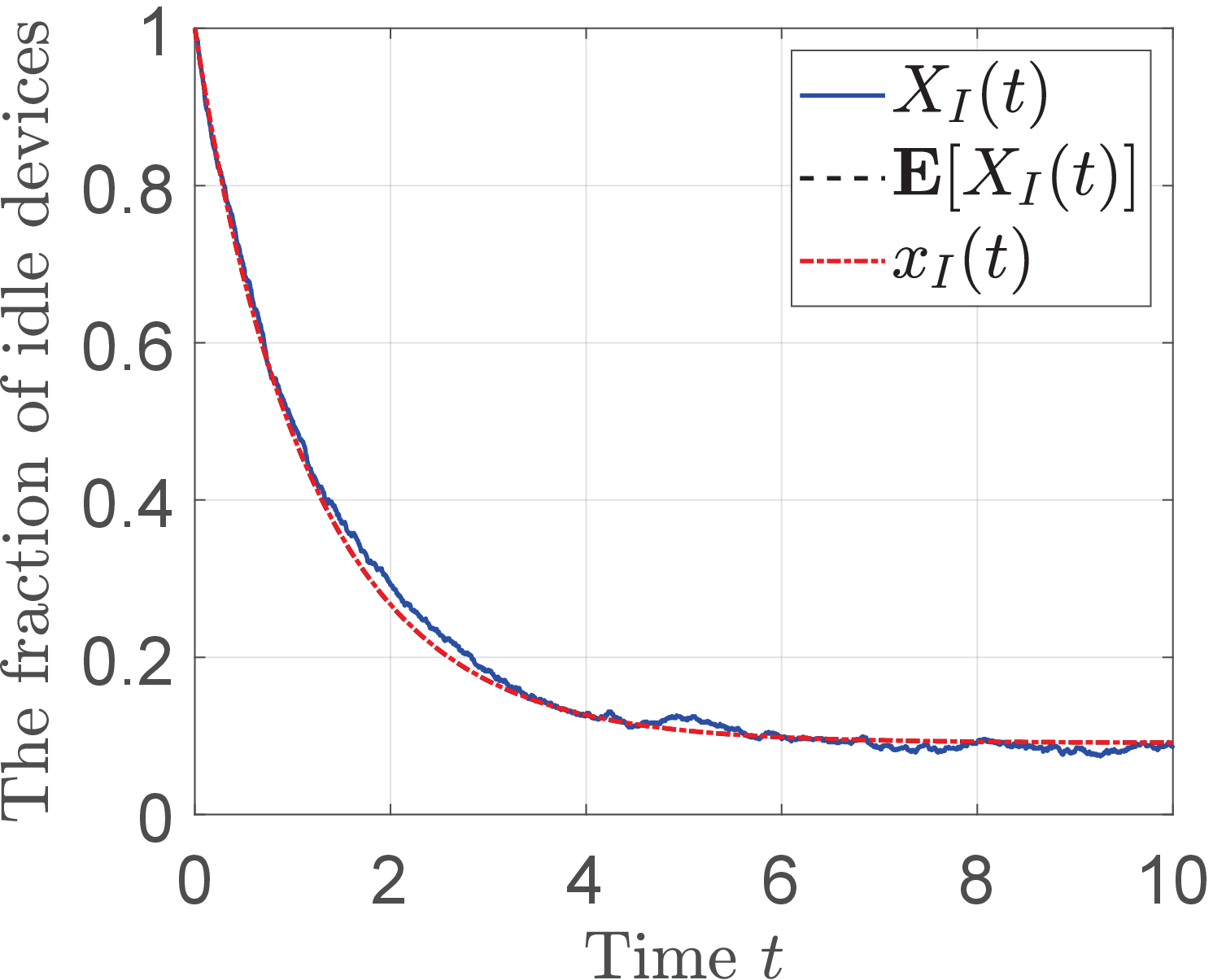}
\subcaption{$N=1000$}\label{fig:N1000}
\end{minipage}
\vspace{-0.1cm}
 \caption{\small{The evolutions of the fraction of IoT devices in state \textit{I} for the CTMC $\bs{X}(t)$ in \eqref{eqn:transition_X} under various numbers of devices $N$. $\lambda=0.8$,  $\mu=1$,  $w=2$, $\gamma=5$, and $p=0.7$.}}
 \label{fig:accurancy}
\vspace{-0.35cm}
\end{figure}

In Fig.~\ref{fig:accurancy}, we illustrate the evolutions of the fraction $X_{\textit{I}}(t)$ of IoT devices in state \textit{I},  with different population sizes of $N=10$, $N=100$, and $N=1000$ of IoT devices, under policy (W).
These results show one simulation trajectory $X_{\textit{I}}(t)$, the average of 10,000 runs of simulation trajectories $\mathbb{E}[X_{\textit{I}}(t)]$ and the mean-field limit $x_{\textit{I}}(t)$ obtained via the ODE in \eqref{eqn:mean-field}. Fig.~\ref{fig:accurancy} clearly shows that the simulation results of one trajectory $X_{\textit{I}}(t)$ are concentrated on the mean-field limit $x_{\textit{I}}(t)$ as $N$ increases. Moreover, we see that the mean-field limit $x_{\textit{I}}(t)$ is very close to the value $\mathbb{E}[X_{\textit{I}}(t)]$, even when $N=10$.
These properties for policy (W) also hold for policies (I) and (S), which can be seen in \cite[Fig. 4]{zhou2020meanfield}.

\begin{figure}[!t]
\begin{centering}
\includegraphics[scale=0.45]{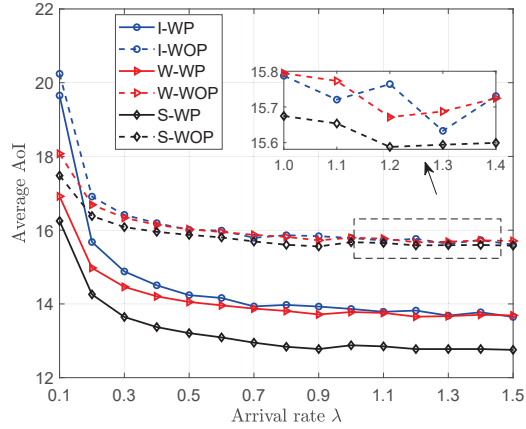}
\vspace{-0.1cm}
 \caption{\small{Average AoI versus the arrival rate in the mean-field limit. $\mu=0.5$, $w=2$, $\gamma=5$, and $p=0.7$.}}
 \label{fig:AoI_lambda_mu05_pe03}
\end{centering}
\vspace{-0.35cm}
\end{figure}

In Fig.~\ref{fig:AoI_lambda_mu05_pe03}, we compare the average AoI for the three policies under different values for the arrival rate $\lambda$ in the mean-field limit. 
\emph{We observe that the average AoI does not necessarily decrease with the arrival rate.} This is because that when $\lambda$ increases, the effective waiting rate $k=w(1-\gamma x_{\textit{S}})$ will decrease  as there would be more devices in state \textit{S}, according to Lemma~\ref{lemma:properties_x}. 
Moreover, we can see that policy (S) always achieves the smallest average AoI compared with policies (I) and (S), and policy (W) can  achieve a smaller average AoI compared with policy (I) only when the channel utilization $\lambda/\mu$ is small.



\begin{figure}[!t]
\centering
\begin{minipage}[h]{.425\linewidth}
\centering
       \includegraphics[scale=0.42]{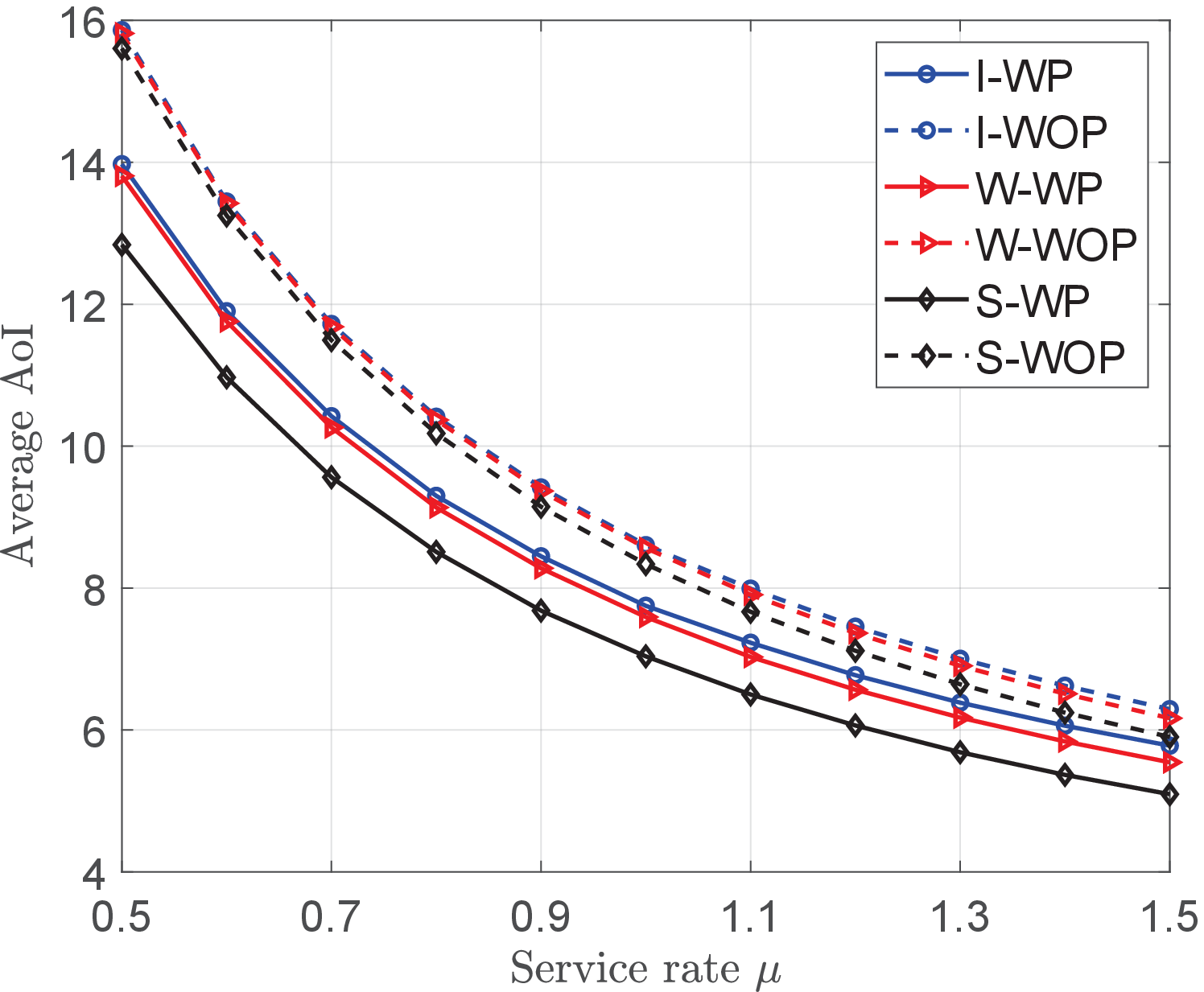}
\subcaption{Service rate $\mu$}\label{fig:AoI_mu}
\end{minipage}
\begin{minipage}[h]{.425\linewidth}
\centering
        \includegraphics[scale=0.42]{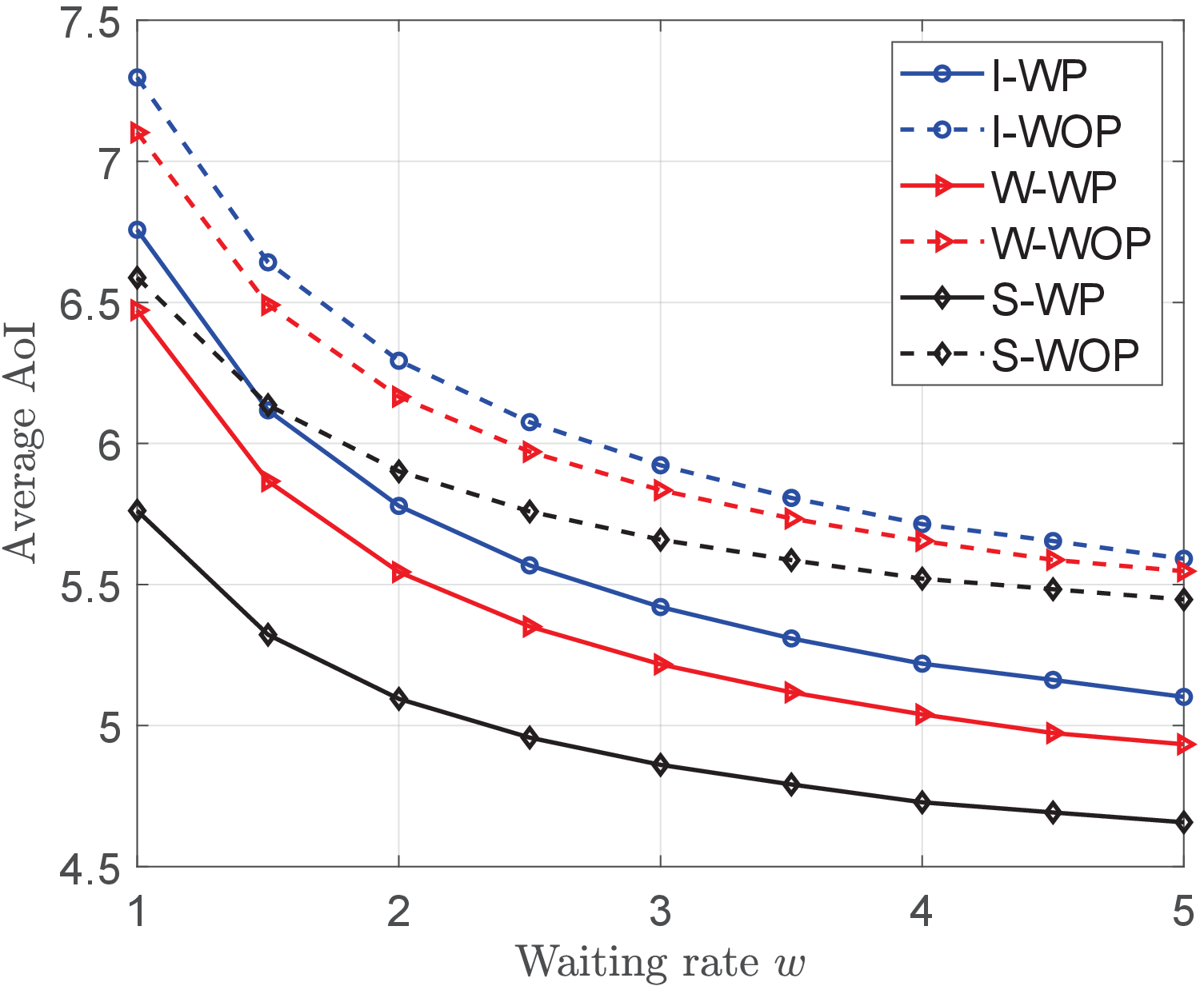}
\subcaption{ Waiting rate $w$}\label{fig:AoI_w}
\end{minipage}

\centering
\begin{minipage}[h]{.425\linewidth}
\centering
       \includegraphics[scale=0.42]{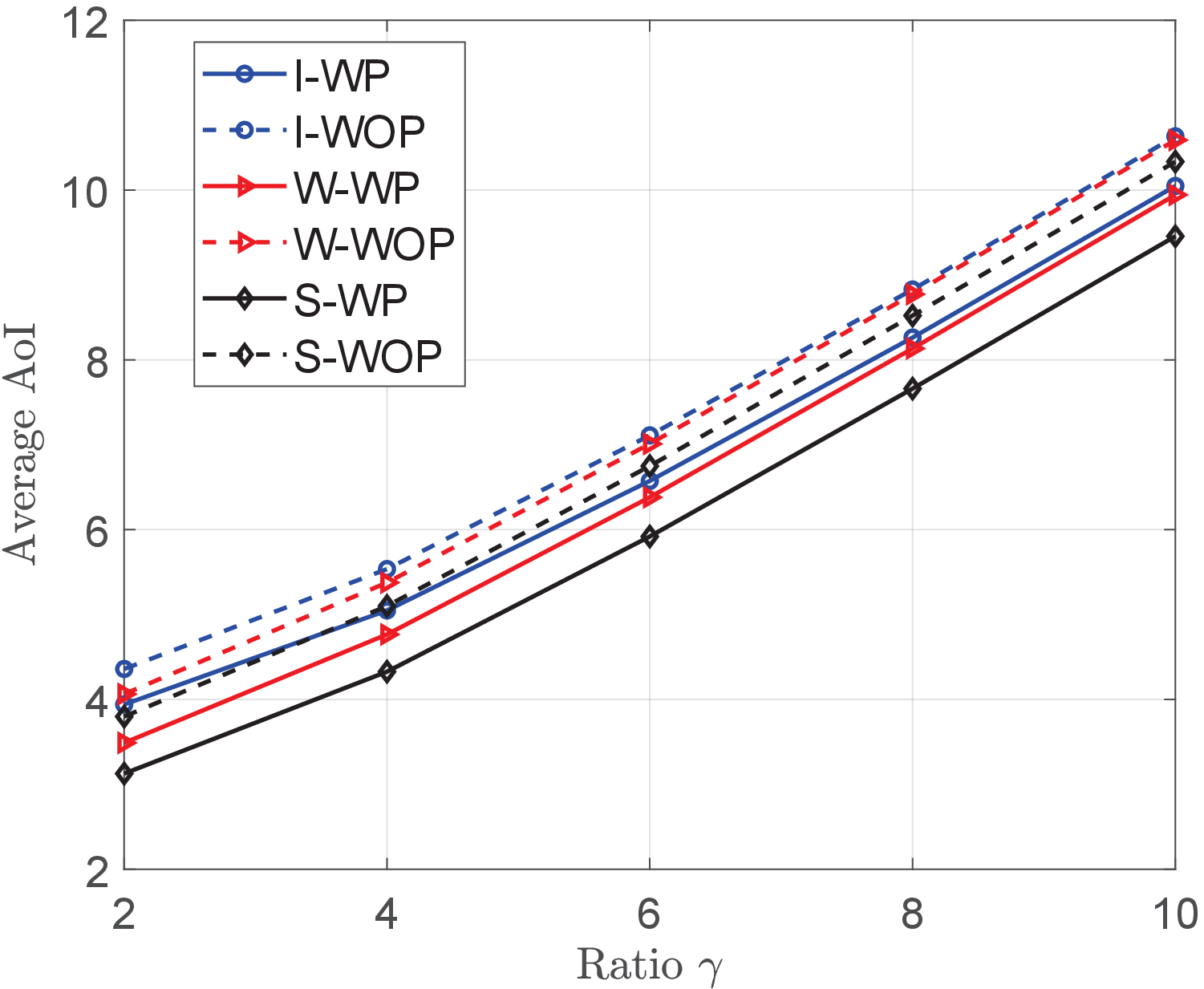}
\subcaption{Ratio $\gamma$}\label{fig:AoI_gamma}
\end{minipage}
\begin{minipage}[h]{.425\linewidth}
\centering
        \includegraphics[scale=0.42]{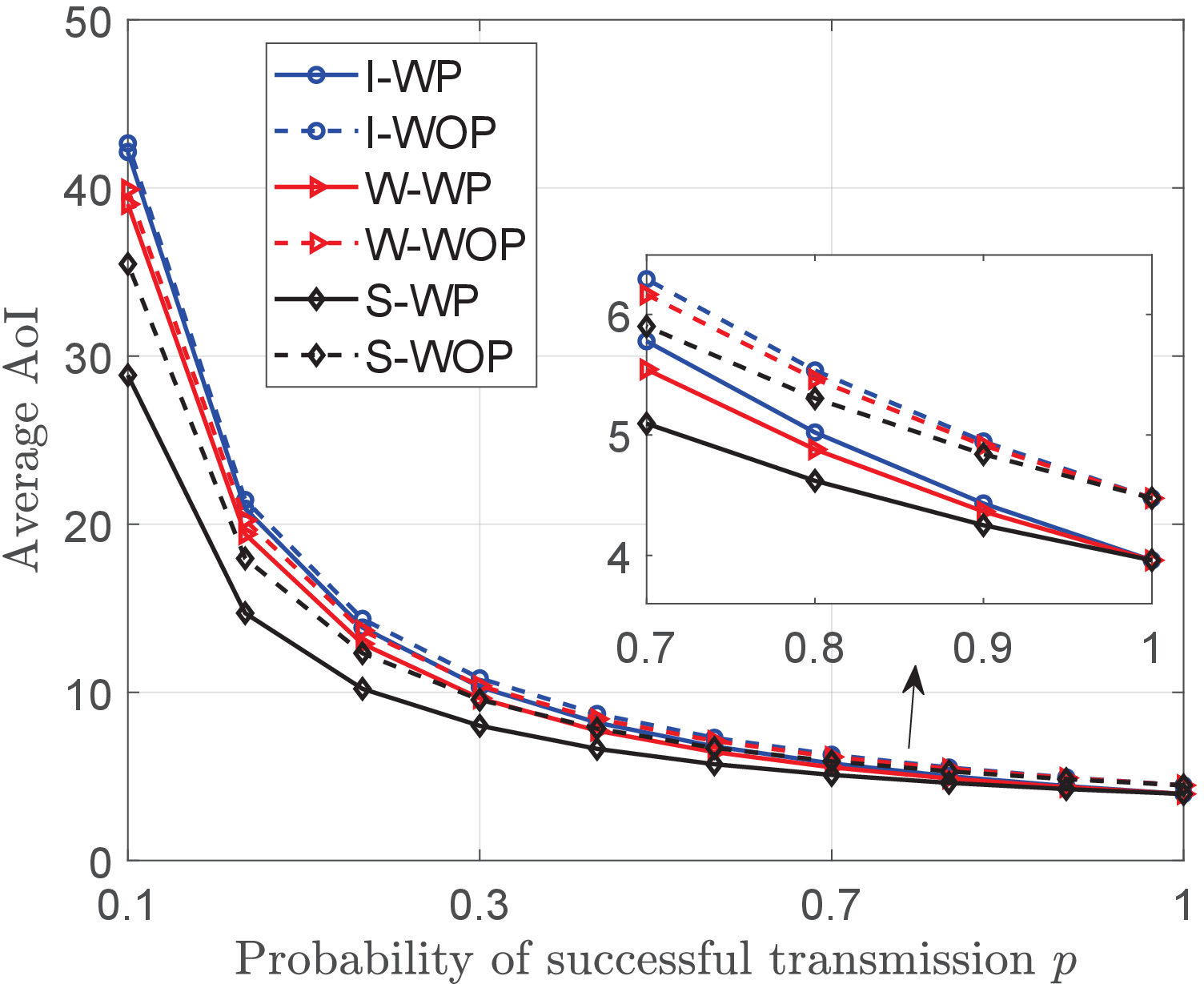}
\subcaption{ Prob. of successful transmission $p$}\label{fig:AoI_p}
\end{minipage}
\vspace{-0.1cm}
 \caption{\small{Average AoI versus system parameters in the mean-field limit. $\lambda=0.8$,  $\mu=1.5$,  $w=2$, $\gamma=5$, and $p=0.7$.}}
 \label{fig:AoI_properties}
\vspace{-0.35cm}
\end{figure}

In Fig.~\ref{fig:AoI_properties}, we further compare the average AoI for the three policies under different values for the service rate $\mu$, the waiting rate $w$, the ratio $\gamma$, and the probability of successful transmission  $p$ in the mean-field limit. 
We can see that, under these three policies with schemes WP and WOP, the average AoI in the mean-field limit decreases with $\mu$, $w$, and $p$, and increases with $\gamma$. This validates the results in Theorem~\ref{theorem:properties-of-aoi}. Moreover, it can be see that policy (S) with scheme WP always achieves the smallest average AoI.

\section{Conclusion}\label{sec:conclusion}
In this paper, we have studied a CSMA-type random access scheme for a ultra-dense IoT monitoring system, under which multiple devices contends for transmitting their status packets to the associated receivers over noisy channels.
We have considered two cases with and without transmission feedback and proposed three policies, i.e., policies (I), (W), and (S).
For each policy with schemes WP and WOP, we have characterized the closed-form expressions of the average AoI of each device for a given stationary distribution of the system, and shown that  scheme WP always achieves a smaller average AoI than scheme WOP.
Then, we have developed a mean-field approximation framework to analyze the asymptotic performance of the considered system in the large population regime. 
We have also studied the effects of the system parameters on the average AoI in the mean-field limit and shown that systems with larger service rates, large waiting rates, more communication resources, and better channel conditions, can achieve better  AoI performance under policies (I) and (S).
Simulation results validate the correctness of the derived closed-form expressions of the average AoI and shown that the mean-filed approximation is very accurate even for a small number of devices.
Moreover, the results show that policy (S) achieves the smallest average AoI than policies (I) and (W), and policy (W) outperforms policy (I) only for systems with low channel utilization. Future works will address key extensions such as investigating the optimization of the system parameters so as to minimize the average AoI under the constraints on the energy cost or the throughput.


\appendices
\section*{Appendix}
\subsection{Proof of Theorem~\ref{theorem:aoi}}\label{app:aoi}
\emph{1) Policy (I):}
We begin with policy (I) under scheme WP by solving for $v_{00}$, $v_{01}$, and $v_{11}$ in \eqref{eqn:vq_I_WP}.
Breaking down \eqref{eqn:vq_I_WP} into its component equations, we obtain:
\begin{subequations}
\begin{align}
&v_{11} = \frac{1}{k+\lambda} \pi_1,\label{eqn:appendix_vq_1}\\
& v_{21} = \frac{1}{\mu + \lambda} (\pi_2 + k \pi_1) = \frac{1}{\mu + \lambda} \left(\pi_2 + \frac{k}{k+\lambda} \pi_1\right),\label{eqn:appendix_vq_2}\\
&\lambda v_{00} = \pi_0 + \mu p v_{21}+  \mu(1-p)v_{20},\label{eqn:appendix_vq_3}\\
&k v_{10} = \pi_1 + \lambda v_{00},\label{eqn:appendix_vq_4}\\
&\mu  v_{20} = \pi_2 + k v_{10}.\label{eqn:appendix_vq_5}
\end{align}
\end{subequations}
By substituting \eqref{eqn:appendix_vq_3} and \eqref{eqn:appendix_vq_4} into \eqref{eqn:appendix_vq_5}, we can find 
\begin{align}
\mu  v_{20} = \pi_0 + \pi_1 + \pi_2 + \mu p v_{21}+  \mu(1-p)v_{20},
\end{align}
from which and by \eqref{eqn:appendix_vq_2}, it follows that
\begin{align}
v_{20} = \frac{1}{\mu}\left(\frac{1}{p} +\frac{\mu}{\mu+\lambda}\left(\pi_2 + \frac{k}{k+\lambda} \pi_1\right)\right).
\end{align}
Then, by \eqref{eqn:appendix_vq_3} and \eqref{eqn:appendix_vq_4}, we have 
\begin{align}
&v_{00} = \frac{1}{\lambda}\left(\pi_0 + \frac{\mu p}{\mu+\lambda}\left(\pi_2 + \frac{k}{k+\lambda} \pi_1\right) + \frac{1-p}{p} +\frac{\mu (1-p)}{\mu+\lambda}\left(\pi_2 + \frac{k}{k+\lambda} \pi_1\right)\right)\nonumber\\
&\quad =  \frac{1}{\lambda}\left(\frac{1}{p}+ \frac{\mu}{\mu+\lambda}\left(\pi_2 + \frac{k}{k+\lambda} \pi_1\right) - \pi_1-\pi_2\right),\\
&v_{10} = \frac{1}{k}\left( \frac{1}{p}+ \frac{\mu}{\mu+\lambda}\left(\pi_2 + \frac{k}{k+\lambda} \pi_1\right)-\pi_2\right).
\end{align}
Thus, according to \eqref{eqn:avg_aoi_yates} and \eqref{eqn:pi_i}, the average AoI under policy (I) with scheme WP is given by
\begin{align}
\bar{\Delta}_{\textrm{I-WP}}&=v_{00}+v_{01} + v_{11}\nonumber\\
&=\left(\frac{1}{\lambda}+\frac{1}{k} + \frac{1}{\mu}\right)\frac{1}{p} + \left(\frac{1}{\lambda}+\frac{1}{k} + \frac{1}{\mu}\right)\frac{\mu}{\mu+\lambda}\left(\pi_2 + \frac{k}{k+\lambda} \pi_1\right) - \frac{1}{\lambda}(\pi_1+\pi_2) -\frac{1}{k} \pi_2 \nonumber\\
&=\left(\frac{1}{\lambda}+\frac{1}{k} + \frac{1}{\mu}\right)\frac{1}{p} + \frac{\lambda + k +\mu}{(\lambda+\mu)(\lambda + k) } - \frac{\lambda + k + \mu}{\lambda k + k\mu + \lambda\mu}.
\end{align}

For policy (I) under scheme WOP, by solving \eqref{eqn:vq_I_WOP} for $v_{00}$, $v_{01}$, and $v_{11}$, we have
\begin{align}
&v_{00} =  \frac{1}{\lambda}\left(\frac{1}{p}+ \pi_2 + \frac{k}{k+\lambda} \pi_1 - \pi_1-\pi_2\right),\\
&v_{10} = \frac{1}{k}\left( \frac{1}{p}+ \pi_2 + \frac{k}{k+\lambda} \pi_1-\pi_2\right),\\
&v_{20} = \frac{1}{\mu}\left(\frac{1}{p} +\pi_2 + \frac{k}{k+\lambda} \pi_1\right).
\end{align}
Then, by  \eqref{eqn:avg_aoi_yates} and  \eqref{eqn:pi_i}, the average AoI under policy (I) with scheme WOP is given by:
\begin{align}
\bar{\Delta}_{\textrm{I-WOP}}&=v_{00}+v_{01} + v_{11}\nonumber\\
&=\left(\frac{1}{\lambda}+\frac{1}{k} + \frac{1}{\mu}\right)\frac{1}{p} + \left(\frac{1}{\lambda}+\frac{1}{k} + \frac{1}{\mu}\right)\left(\pi_2 + \frac{k}{k+\lambda} \pi_1\right) - \frac{1}{\lambda}(\pi_1+\pi_2) -\frac{1}{k} \pi_2 \nonumber\\
&=\left(\frac{1}{\lambda}+\frac{1}{k} + \frac{1}{\mu}\right)\frac{1}{p} + \frac{1}{\mu} + \frac{1}{\lambda +k} - \frac{\lambda + k + \mu}{\lambda k + k\mu + \lambda\mu}.
\end{align}
We complete the proof for policy (I).

\begin{figure}[!t]
\centering
\begin{minipage}[h]{.49\linewidth}{}
\centering
\begin{tikzpicture}[scale=0.15]
\tikzstyle{every node}+=[inner sep=0pt]
\draw [thick,black]  (22.5,-43.8) circle (3);
\draw (22.5,-43.8) node {$0$};
\draw [thick,black]  (30.8,-31.8) circle (3);
\draw (30.8,-31.8) node {$1$};
\draw [thick,black]  (39.9,-43.8) circle (3);
\draw (39.9,-43.8) node {$2$};
\draw [thick,black]  (24.21,-41.33) -- (29.09,-34.27);
\fill [thick,black]  (29.09,-34.27) -- (28.23,-34.64) -- (29.05,-35.21);
\draw (26.05,-36.44) node [left] {$1$};
\draw [thick,black]  (33.555,-32.969) arc (60.19073:14.15778:12.618);
\fill [thick,black]  (39.52,-40.83) -- (39.81,-39.93) -- (38.84,-40.18);
\draw (37.91,-34.89) node [right] {$2$};
\draw [thick,black]  (27.938,-32.66) arc (314.46724:26.46724:2.25);
\draw (23.43,-30.45) node [left] {$3$};
\fill [thick,black]  (28.38,-30.05) -- (28.17,-29.13) -- (27.46,-29.83);
\draw [thick,black]  (42.58,-42.477) arc (144:-144:2.25);
\draw (47.15,-43.8) node [right] {$6$};
\fill [thick,black]  (42.58,-45.12) -- (42.93,-46) -- (43.52,-45.19);
\draw [thick,black]  (36.9,-43.8) -- (25.5,-43.8);
\fill [thick,black]  (25.5,-43.8) -- (26.3,-44.3) -- (26.3,-43.3);
\draw (31.2,-43.3) node [above] {$4$};
\draw [thick,black]  (37.187,-42.535) arc (-121.42402:-164.22747:13.38);
\fill [thick,black]  (31.29,-34.75) -- (31.02,-35.66) -- (31.98,-35.39);
\draw (32.93,-40.6) node [left] {$5$};
\end{tikzpicture}
 \subcaption{Scheme WP.}\label{fig:shs_ctmc_W_wp}
\end{minipage}
\begin{minipage}[h]{.49\linewidth}
\centering
\begin{tikzpicture}[scale=0.15]
\tikzstyle{every node}+=[inner sep=0pt]
\draw [thick,black]  (22.5,-43.8) circle (3);
\draw (22.5,-43.8) node {$0$};
\draw [thick,black]  (30.8,-31.8) circle (3);
\draw (30.8,-31.8) node {$1$};
\draw [thick,black]  (39.9,-43.8) circle (3);
\draw (39.9,-43.8) node {$2$};
\draw [thick,black]  (24.21,-41.33) -- (29.09,-34.27);
\fill [thick,black]  (29.09,-34.27) -- (28.23,-34.64) -- (29.05,-35.21);
\draw (26.05,-36.44) node [left] {$1$};
\draw [thick,black]  (33.555,-32.969) arc (60.19073:14.15778:12.618);
\fill [thick,black]  (39.52,-40.83) -- (39.81,-39.93) -- (38.84,-40.18);
\draw (37.91,-34.89) node [right] {$2$};
\draw [thick,black]  (27.983,-32.796) arc (317.20662:29.20662:2.25);
\draw (23.43,-30.88) node [left] {$3$};
\fill [thick,black]  (28.3,-30.17) -- (28.05,-29.26) -- (27.37,-29.99);
\draw [thick,black]  (36.9,-43.8) -- (25.5,-43.8);
\fill [thick,black]  (25.5,-43.8) -- (26.3,-44.3) -- (26.3,-43.3);
\draw (31.2,-43.3) node [above] {$4$};
\draw [thick,black]  (37.187,-42.535) arc (-121.42402:-164.22747:13.38);
\fill [thick,black]  (31.29,-34.75) -- (31.02,-35.66) -- (31.98,-35.39);
\draw (32.93,-40.6) node [left] {$5$};
\end{tikzpicture}
 \subcaption{Scheme WOP.}\label{fig:shs_ctmc_W_wop}
\end{minipage}
\vspace{-0.1cm}
 \caption{\small{Illustration of the SHS Markov chain under policy (W).}}
 \label{fig:shs-ctmc-W}
\vspace{-0.35cm}
\end{figure}
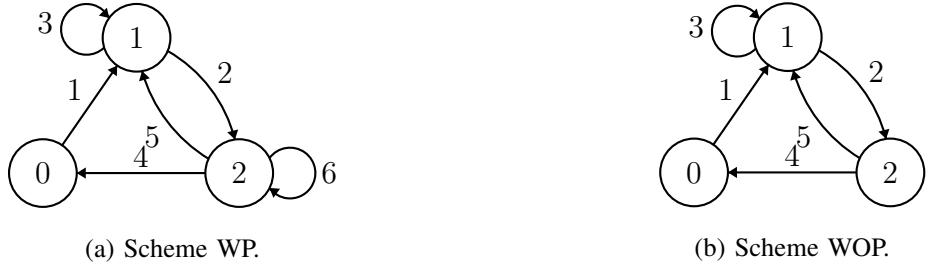

\emph{2) Policy (W):}
Based on the SHS approach in a similar manner for policy (I), we now derive the average AoI under policy (W) with schemes WP and WOP. For policy (W) with scheme WP, the SHS Markov chain is illustrated in Fig.~\ref{fig:shs_ctmc_W_wp} and the corresponding transitions are summarized in Table~\ref{table:shs_ctms_W_wp}. 
It can be seen that, the transitions $l=1,2,3,4,6$ in Table~\ref{table:shs_ctms_W_wp} are the same to those in Table~\ref{table:shs_ctms_I_wp}.
For transition $l=5$, the device completes its service and fails to deliver a status packet to the receiver. Thus, the device will transition from state $2$ to state $1$ with the undelivered packet under policy (W), and the AoI at the receiver and the age of the packet at the device remain unchanged, i.e., $z_0'=z_0$ and $z_1'=z_1$.
Moreover, the evolution of the age-related continuous state $\bs{z}(t)$ depends on the discrete state $q(t)$ with the same rates $\bs{b}_q$ in \eqref{eqn:bq}. 

From Fig.~\ref{fig:shs_ctmc_W_wp} and Table~\ref{table:shs_ctms_W_wp}, the stationary distribution $\bs{\pi}=[\pi_0~ \pi_1 ~\pi_2]$ of $q(t)$ satisfies:
\begin{align*}
&\lambda \pi_0 =  \mu p\pi_2,~k \pi_1 =  \lambda \pi_0 + \mu(1-p) \pi_2,\\
&\mu \pi_2 =  k \pi_1,~\pi_0 + \pi_1 + \pi_2 = 1.
\end{align*}
Then, we have:
\begin{align}\label{eqn:pi_w}
\bs{\pi}=[\pi_0~ \pi_1 ~\pi_2] = \frac{1}{\lambda k + \lambda\mu + k\mu p}\left[ k\mu p ~ \lambda\mu ~ k \lambda\right].
\end{align}
By \eqref{eqn:v_vector_yates} and Table~\ref{table:shs_ctms_W_wp}, we have
\begin{subequations}\label{eqn:vq_W_WP}
\begin{align}
&\lambda [v_{00}~v_{01}] = \pi_0[1~0] + \mu p [v_{21}~0],\\
&(k+\lambda) [v_{10}~v_{11}] = \pi_1[1~1]  + \lambda [v_{00}~0] + \lambda[v_{10}~0]+ \mu (1-p) [v_{20}~v_{21}],\\
&(\mu+\lambda) [v_{20}~v_{21}] = \pi_2[1~1]  + k [v_{10}~v_{11}] + \lambda[v_{20}~0] .
\end{align}
\end{subequations}
Solving \eqref{eqn:vq_W_WP} for $v_{00}$, $v_{01}$, and $v_{11}$, we obtain 
\begin{align}
&v_{00} =  \frac{p}{\lambda}\left(\frac{1}{p}\pi_0+ \frac{\mu (k+\lambda)}{(\lambda+\mu)(k+\lambda)-k\mu (1-p)}\left(\pi_2 + \frac{k}{k+\lambda} \pi_1\right)\right),\\
&v_{10} = \frac{1}{k}\left(\frac{1}{p} (\pi_0 +  \pi_1) + \frac{\mu (k+\lambda)}{(\lambda+\mu)(k+\lambda)-k\mu (1-p)}\left(\pi_2 + \frac{k}{k+\lambda} \pi_1\right)\right),\\
&v_{20} = \frac{1}{\mu}\left(\frac{1}{p} +\frac{\mu (k+\lambda)}{(\lambda+\mu)(k+\lambda)-k\mu (1-p)}\left(\pi_2 + \frac{k}{k+\lambda} \pi_1\right)\right).
\end{align}
Then, by \eqref{eqn:avg_aoi_yates} and \eqref{eqn:pi_w}, the average AoI for policy (W) with scheme WP is given by
\begin{align}
&\bar{\Delta}_{\textrm{W-WP}}=v_{00}+v_{01} + v_{11}\nonumber\\
& = \left(\frac{p}{\lambda} + \frac{1}{k} + \frac{1}{\mu}\right)\left(\frac{1}{p}  +  \frac{\mu (k+\lambda)}{(\lambda+\mu)(k+\lambda)-k\mu (1-p)}\left(\pi_2 + \frac{k}{k+\lambda} \pi_1\right)\right) - \frac{1}{\lambda}(\pi_1+\pi_2) - \frac{1}{k}\pi_2\nonumber\\
& = \left(\frac{p}{\lambda} + \frac{1}{k} + \frac{1}{\mu}\right) \left(\frac{1}{p}  + \frac{\mu (k+\lambda)}{(\lambda+\mu)(k+\lambda)-k\mu (1-p)}\left(\frac{k+\lambda+\mu}{k+\lambda}\right)\pi_2\right) - \left(\frac{k+\lambda+\mu}{k \lambda}\right)\pi_2\nonumber\\
& = \frac{1}{p} \left(\frac{p}{\lambda} + \frac{1}{k} + \frac{1}{\mu}\right)+ \frac{\lambda + k + \mu}{(\lambda+\mu)(k+\lambda)-k\mu (1-p)}- \frac{\lambda + k + \mu}{\lambda k + \lambda\mu + k\mu p}.
\end{align}

For policy (W) with scheme WOP, the analysis is similar. However, in that case,  there is no transition $l=6$ for the corresponding SHS Markov chain (as shown in Fig.~\ref{fig:shs_ctmc_W_wop}), because preemption in service is not allowed. The transitions $l=1,2\cdots,5$ are the same to those in Table~\ref{table:shs_ctms_W_wp} and the stationary distribution is the same to that in \eqref{eqn:pi_w}. 
By \eqref{eqn:v_vector_yates} and Table~\ref{table:shs_ctms_W_wp}, we have
\begin{subequations}\label{eqn:vq_W_WOP}
\begin{align}
&\lambda [v_{00}~v_{01}] = \pi_0[1~0] + \mu p [v_{21}~0],\\
&(k+\lambda) [v_{10}~v_{11}] = \pi_1[1~1]  + \lambda [v_{00}~0] + \lambda[v_{10}~0]+ \mu (1-p) [v_{20}~v_{21}],\\
&(\mu+\lambda) [v_{20}~v_{21}] = \pi_2[1~1]  + k [v_{10}~v_{11}].
\end{align}
\end{subequations}
Solving \eqref{eqn:vq_W_WOP} for $v_{00}$, $v_{01}$, and $v_{11}$, we have
\begin{align}
&v_{00} =  \frac{p}{\lambda}\left(\frac{1}{p}\pi_0+ \pi_2 + \frac{k}{k p +\lambda}(\pi_1 + (1-p)\pi_2)\right),\\
&v_{10} = \frac{1}{k}\left(\frac{1}{p} + \frac{k}{k p +\lambda}(\pi_1 + (1-p)\pi_2)\right),\\
&v_{20} = \frac{1}{\mu}\left(\frac{1}{p}+ \pi_2 + \frac{k}{k p +\lambda}(\pi_1 + (1-p)\pi_2)\right).
\end{align}
Then, the average AoI for policy (W) with scheme WOP is given by
\begin{align}
\bar{\Delta}_{\textrm{W-WOP}}=v_{00}+v_{01} + v_{11}
& = \frac{1}{p} \left(\frac{p}{\lambda} + \frac{1}{k} + \frac{1}{\mu}\right) + \frac{1}{\mu}\frac{k\pi_1 + (k+\lambda)\pi_2}{kp + \lambda}\nonumber\\
&= \frac{1}{p}\left(\frac{p}{\lambda} + \frac{1}{k} + \frac{1}{\mu}\right) + \frac{\lambda + k + \mu}{\mu(kp+\lambda)}  - \frac{\lambda + k + \mu}{\lambda k + \lambda\mu + k\mu p}.
\end{align}
We complete the proof for policy (I).

\begin{table}[!t]
\centering
\caption{\small{Transitions for the SHS Markov chain under policy (W) with scheme WP in Fig.~\ref{fig:shs_ctmc_W_wp}.}}\label{table:shs_ctms_W_wp}
\begin{tabular}{cccccc}
\hline
$l$ & $q_l\to q_l'$ & $\lambda^{l}$  & $\bs{z}\bs{A}_l$ & $\bs{A}_l$  & $\bs{v}_{q_l} \bs{A}_l$ \\
\hline
$1$ & $0 \to 1$     & $\lambda$   & $\begin{bmatrix} z_0 & 0 \end{bmatrix}$ & $\begin{bmatrix} 1 & 0 \\ 0 & 0 \end{bmatrix}$ & $\begin{bmatrix} \bar{v}_{00} & 0 \end{bmatrix}$    \\         
$2$ & $1 \to 2$     & $k$   & $\begin{bmatrix} z_0 & z_1 \end{bmatrix}$ & $\begin{bmatrix} 1 & 0 \\ 0 & 1 \end{bmatrix}$ & $\begin{bmatrix} \bar{v}_{10} & \bar{v}_{11} \end{bmatrix}$    \\  
$3$ & $1 \to 1$     & $\lambda$     & $\begin{bmatrix} z_0 & 0 \end{bmatrix}$ & $\begin{bmatrix} 1 & 0 \\ 0 & 0 \end{bmatrix}$ & $\begin{bmatrix} \bar{v}_{10} & 0 \end{bmatrix}$    \\         
$4$ & $2 \to 0$     & $\mu p$ & $\begin{bmatrix} z_1 & 0 \end{bmatrix}$ & $\begin{bmatrix} 0 & 0 \\ 1 & 0 \end{bmatrix}$ & $\begin{bmatrix} \bar{v}_{21} & 0 \end{bmatrix}$    \\         
$5$ & $2 \to 1$     & $\mu (1-p)$   & $\begin{bmatrix} z_0 & z_1 \end{bmatrix}$ & $\begin{bmatrix} 1 & 0 \\ 0 & 1 \end{bmatrix}$ & $\begin{bmatrix} \bar{v}_{20} & \bar{v}_{21} \end{bmatrix}$    \\         
$6$ & $2 \to 2$     & $\lambda$  & $\begin{bmatrix} z_0 & 0 \end{bmatrix}$ & $\begin{bmatrix} 1 & 0 \\ 0 & 0 \end{bmatrix}$ & $\begin{bmatrix} \bar{v}_{20} & 0 \end{bmatrix}$     \\
\hline 
\end{tabular}
\vspace{-0.35cm}
\end{table}

\begin{figure}[!t]
\centering
\begin{minipage}[h]{.49\linewidth}{}
\centering
\begin{tikzpicture}[scale=0.15]
\tikzstyle{every node}+=[inner sep=0pt]
\draw [thick,black] (22.5,-43.8) circle (3);
\draw (22.5,-43.8) node {$0$};
\draw [thick,black] (30.8,-31.8) circle (3);
\draw (30.8,-31.8) node {$1$};
\draw [thick,black] (39.8,-43.8) circle (3);
\draw (39.8,-43.8) node {$2$};
\draw [thick,black] (24.21,-41.33) -- (29.09,-34.27);
\fill [thick,black] (29.09,-34.27) -- (28.23,-34.64) -- (29.05,-35.21);
\draw (26.05,-36.44) node [left] {$1$};
\draw [thick,black] (27.972,-32.766) arc (316.59891:28.59891:2.25);
\draw (23.43,-30.79) node [left] {$3$};
\fill [thick,black] (28.31,-30.14) -- (28.08,-29.23) -- (27.39,-29.96);
\draw [thick,black] (36.8,-43.8) -- (25.5,-43.8);
\fill [thick,black] (25.5,-43.8) -- (26.3,-44.3) -- (26.3,-43.3);
\draw (31.15,-43.3) node [above] {$4$};
\draw [thick,black] (32.6,-34.2) -- (38,-41.4);
\fill [thick,black] (38,-41.4) -- (37.92,-40.46) -- (37.12,-41.06);
\draw (35.88,-36.4) node [right] {$2$};
\draw [thick,black] (38.892,-40.953) arc (225.42739:-62.57261:2.25);
\draw (41.18,-36.53) node [above] {$5$};
\fill [thick,black] (41.51,-41.35) -- (42.42,-41.13) -- (41.71,-40.43);
\draw [thick,black] (42.48,-42.477) arc (144:-144:2.25);
\draw (47.05,-43.8) node [right] {$6$};
\fill [thick,black] (42.48,-45.12) -- (42.83,-46) -- (43.42,-45.19);
\end{tikzpicture}
 \subcaption{Scheme WP.}\label{fig:shs_ctmc_S_wp}
\end{minipage}
\begin{minipage}[h]{.49\linewidth}
\centering
\begin{tikzpicture}[scale=0.15]
\tikzstyle{every node}+=[inner sep=0pt]
\draw [thick,black] (22.5,-43.8) circle (3);
\draw (22.5,-43.8) node {$0$};
\draw [thick,black] (30.8,-31.8) circle (3);
\draw (30.8,-31.8) node {$1$};
\draw [thick,black] (39.8,-43.8) circle (3);
\draw (39.8,-43.8) node {$2$};
\draw [thick,black] (24.21,-41.33) -- (29.09,-34.27);
\fill [thick,black] (29.09,-34.27) -- (28.23,-34.64) -- (29.05,-35.21);
\draw (26.05,-36.44) node [left] {$1$};
\draw [thick,black] (27.972,-32.766) arc (316.59891:28.59891:2.25);
\draw (23.43,-30.79) node [left] {$3$};
\fill [thick,black] (28.31,-30.14) -- (28.08,-29.23) -- (27.39,-29.96);
\draw [thick,black] (36.8,-43.8) -- (25.5,-43.8);
\fill [thick,black] (25.5,-43.8) -- (26.3,-44.3) -- (26.3,-43.3);
\draw (31.15,-43.3) node [above] {$4$};
\draw [thick,black] (32.6,-34.2) -- (38,-41.4);
\fill [thick,black] (38,-41.4) -- (37.92,-40.46) -- (37.12,-41.06);
\draw (35.88,-36.4) node [right] {$2$};
\draw [thick,black] (38.892,-40.953) arc (225.42739:-62.57261:2.25);
\draw (41.18,-36.53) node [above] {$5$};
\fill [thick,black] (41.51,-41.35) -- (42.42,-41.13) -- (41.71,-40.43);
\end{tikzpicture}
 \subcaption{Scheme WOP.}\label{fig:shs_ctmc_S_wop}
\end{minipage}
\vspace{-0.1cm}
 \caption{\small{Illustration of the SHS Markov chain under policy (S).}}
 \label{fig:shs-ctmc-S}
\vspace{-0.35cm}
\end{figure}
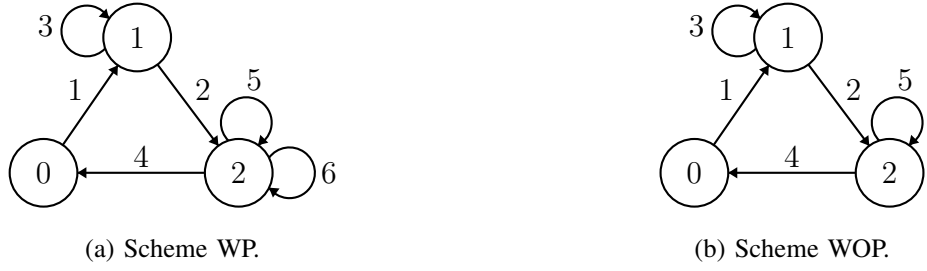

\emph{3) Policy (S):}
Similarly, we apply the SHS method to derive the average AoI for policy (S) with schemes WP and WOP. 
We illustrate the SHS Markov chain for policy (S) with scheme WP in Fig.~\ref{fig:shs_ctmc_S_wp} and the corresponding transitions in Table~\ref{table:shs_ctms_S_wp}.
We can see that, the transitions $l=1,2,3,4,6$ in Table~\ref{table:shs_ctms_S_wp} are the same to those in Table~\ref{table:shs_ctms_I_wp}.
For transition $l=5$, the device completes its service and fails to deliver a status packet to the receiver. Thus, the device will stay at state $2$ to attempt another transmission with the undelivered packet under policy (S), and the AoI at the receiver and the age of the packet at the device remain unchanged, i.e., $z_0'=z_0$ and $z_1'=z_1$.
The evolution of the continuous state $\bs{z}(t)$ also depends on the discrete state $q(t)$ with the same rates $\bs{b}_q$ in \eqref{eqn:bq}. 
From Fig.~\ref{fig:shs_ctmc_S_wp} and Table~\ref{table:shs_ctms_S_wp}, the stationary distribution $\bs{\pi}=[\pi_0~ \pi_1 ~\pi_2]$ of $q(t)$ satisfies the following system of linear equations:
\begin{align*}
&\lambda \pi_0 = k \pi_1 = \mu p \pi_2,~\pi_0 + \pi_1 + \pi_2 = 1.
\end{align*}
Then, we have:
\begin{align}\label{eqn:pi_S}
\bs{\pi}=[\pi_0~ \pi_1 ~\pi_2] = \frac{1}{\lambda k + (k + \lambda)\mu p}\left[ k\mu p ~ \lambda\mu p ~ k \lambda\right].
\end{align}
By \eqref{eqn:v_vector_yates} and Table~\ref{table:shs_ctms_W_wp}, we have
\begin{subequations}\label{eqn:vq_S_WP}
\begin{align}
&\lambda [v_{00}~v_{01}] = \pi_0[1~0] + \mu p [v_{21}~0],\\
&(k+\lambda) [v_{10}~v_{11}] = \pi_1[1~1]  + \lambda [v_{00}~0] + \lambda[v_{10}~0],\\
&(\mu+\lambda) [v_{20}~v_{21}] = \pi_2[1~1]  + k [v_{10}~v_{11}] + \lambda[v_{20}~0] + \mu (1-p) [v_{20}~v_{21}].
\end{align}
\end{subequations}
Solving \eqref{eqn:vq_S_WP} for $v_{00}$, $v_{01}$, and $v_{11}$, we have 
\begin{align}
&v_{00} =  \frac{1}{\lambda}\left(\pi_0+ \frac{\mu p}{\mu p +\lambda}\left(\pi_2 + \frac{k}{k+\lambda} \pi_1\right)\right),\\
&v_{10} = \frac{1}{k}\left(\pi_0 +  \pi_1 + \frac{\mu p}{\mu p +\lambda}\left(\pi_2 + \frac{k}{k+\lambda} \pi_1\right)\right),\\
&v_{20} = \frac{1}{\mu p}\left(1 +\frac{\mu p}{\mu p +\lambda}\left(\pi_2 + \frac{k}{k+\lambda} \pi_1\right)\right).
\end{align}
Then, by \eqref{eqn:avg_aoi_yates} and \eqref{eqn:pi_S}, the average AoI for policy (S) with scheme WP is given by
\begin{align}
\bar{\Delta}_{\textrm{S-WP}}&=v_{00}+v_{01} + v_{11}\nonumber\\
& = \left(\frac{1}{\lambda} + \frac{1}{k} + \frac{1}{\mu p}\right)\left(1+ \frac{\mu p}{\mu p +\lambda}\left(\pi_2 + \frac{k}{k+\lambda} \pi_1\right)\right) - \frac{1}{\lambda} (\pi_1 +\pi_2) -\frac{1}{k} \pi_2   \nonumber\\
& = \frac{1}{\lambda} + \frac{1}{k} + \frac{1}{\mu p} + \frac{\mu p + k + \lambda}{(\lambda + \mu p)(\lambda + k)} - \frac{\lambda + k + \mu p}{\lambda k + (k + \lambda)\mu p}.
\end{align}

\begin{table}[!t]
\centering
\caption{\small{Transitions for the SHS Markov chain under policy (S) with scheme WP in Fig.~\ref{fig:shs_ctmc_S_wp}.}}\label{table:shs_ctms_S_wp}
\begin{tabular}{cccccc}
\hline
$l$ & $q_l\to q_l'$ & $\lambda^{l}$  & $\bs{z}\bs{A}_l$ & $\bs{A}_l$  & $\bs{v}_{q_l} \bs{A}_l$ \\
\hline
$1$ & $0 \to 1$     & $\lambda$   & $\begin{bmatrix} z_0 & 0 \end{bmatrix}$ & $\begin{bmatrix} 1 & 0 \\ 0 & 0 \end{bmatrix}$ & $\begin{bmatrix} \bar{v}_{00} & 0 \end{bmatrix}$    \\         
$2$ & $1 \to 2$     & $k$   & $\begin{bmatrix} z_0 & z_1 \end{bmatrix}$ & $\begin{bmatrix} 1 & 0 \\ 0 & 1 \end{bmatrix}$ & $\begin{bmatrix} \bar{v}_{10} & \bar{v}_{11} \end{bmatrix}$    \\  
$3$ & $1 \to 1$     & $\lambda$     & $\begin{bmatrix} z_0 & 0 \end{bmatrix}$ & $\begin{bmatrix} 1 & 0 \\ 0 & 0 \end{bmatrix}$ & $\begin{bmatrix} \bar{v}_{10} & 0 \end{bmatrix}$    \\         
$4$ & $2 \to 0$     & $\mu p$ & $\begin{bmatrix} z_1 & 0 \end{bmatrix}$ & $\begin{bmatrix} 0 & 0 \\ 1 & 0 \end{bmatrix}$ & $\begin{bmatrix} \bar{v}_{21} & 0 \end{bmatrix}$    \\         
$5$ & $2 \to 2$     & $\mu (1-p)$   & $\begin{bmatrix} z_0 & z_1 \end{bmatrix}$ & $\begin{bmatrix} 1 & 0 \\ 0 & 1 \end{bmatrix}$ & $\begin{bmatrix} \bar{v}_{20} & \bar{v}_{21} \end{bmatrix}$    \\         
$6$ & $2 \to 2$     & $\lambda$  & $\begin{bmatrix} z_0 & 0 \end{bmatrix}$ & $\begin{bmatrix} 1 & 0 \\ 0 & 0 \end{bmatrix}$ & $\begin{bmatrix} \bar{v}_{20} & 0 \end{bmatrix}$     \\
\hline 
\end{tabular}
\vspace{-0.35cm}
\end{table}

For policy (S) with scheme WOP, as illustrated in Fig.~\ref{fig:shs_ctmc_S_wop}, there is no transition $l=6$ at state $2$ in the SHS Markov chain.
The transitions $l=1,2\cdots,5$ are the same to those in Table~\ref{table:shs_ctms_S_wp} and the stationary distribution is the same to that in \eqref{eqn:pi_S}. Similarly, we have 
\begin{subequations}\label{eqn:vq_S_WOP}
\begin{align}
&\lambda [v_{00}~v_{01}] = \pi_0[1~0] + \mu p [v_{21}~0],\\
&(k+\lambda) [v_{10}~v_{11}] = \pi_1[1~1]  + \lambda [v_{00}~0] + \lambda[v_{10}~0],\\
&(\mu+\lambda) [v_{20}~v_{21}] = \pi_2[1~1]  + k [v_{10}~v_{11}]  + \mu (1-p) [v_{20}~v_{21}],
\end{align}
\end{subequations}
from which, it follows that
\begin{align}
&v_{00} =  \frac{1}{\lambda}\left(\pi_0+ \pi_2 + \frac{k}{k+\lambda} \pi_1\right),\\
&v_{10} = \frac{1}{k}\left(1 + \frac{k}{k+\lambda} \pi_1\right),\\
&v_{20} = \frac{1}{\mu p}\left(1 +\pi_2 + \frac{k}{k+\lambda} \pi_1\right).
\end{align}
Thus, the average AoI under policy (S) with scheme WOP is given by
\begin{align}
\bar{\Delta}_{\textrm{S-WOP}}=v_{00}+v_{01} + v_{11}
& = \frac{1}{\lambda} + \frac{1}{k} + \frac{2}{\mu p} +\left(\frac{1}{k+\lambda} + \frac{1}{\mu p}\right) \pi_2   \nonumber\\
& = \frac{1}{\lambda} + \frac{1}{k} + \frac{2}{\mu p} + \frac{1}{\lambda + k } - \frac{\lambda + k + \mu p}{\lambda k + (k + \lambda)\mu p}.
\end{align}
We complete the proof of Theorem~\ref{theorem:aoi}.

\subsection{Proof of Theorem~\ref{theorem:mean-field}}\label{app:mean-field}
First, we show that the equilibrium point $\bs{x}^*$ of the mean-field model in \eqref{eqn:mean-field} is unique and is given in \eqref{eqn:mean-field-solution}.
By \eqref{eqn:mean-field}, and by the definition that $x_I^*+x_W^*+x_S^*=1$, we can see that $\bs{x}^*$ satisfies the following fixed point equations:
\begin{subequations}\label{eqn:mean-field-solution-proof}
\begin{align}
&x_{\textit{I}}^* = \frac{\mu p}{\lambda}x_{\textit{S}}^*,~x_{\textit{W}}^* = \frac{\mu x_{\textit{S}}^*}{w(1-\gamma x_{\textit{S}}^*)},\label{eqn:xixw}\\
&x_{\textit{S}}^* = \frac{1/\mu}{p/\lambda+1/(w(1-\gamma x_{\textit{S}}^*))+1/\mu} = \frac{\lambda w (1-\gamma x_{\textit{S}}^*)}{(\lambda+\mu p)w (1-\gamma x_{\textit{S}}^*) + \lambda\mu}.\label{eqn:xs}
       \end{align}
\end{subequations}
Then, to show the uniqueness of $\bs{x}^*$, it is sufficient to show that the solution $x_{\textit{S}}^*$ to \eqref{eqn:xs} is unique. 
To see this, we transform \eqref{eqn:xs} into the following quadratic equation:
\begin{align}\label{eqn:solving_x_s}
w(\lambda+\mu p)\gamma (x_{\textit{S}}^*)^2 - (w(\lambda+\mu p+\lambda\gamma)+\lambda\mu)x_{\textit{S}}^* + \lambda w=0.
\end{align}
Note that, under  policy (W), the probability that any channel is sensed busy is $\gamma x_{\textit{S}}^*$, and thus we have $0\leq x_{\textit{S}}^*< \frac{1}{\gamma}$. By following the similar approach for \cite[Theorem 2]{zhou2020meanfield}, we can show that there is only one feasible solution of \eqref{eqn:xs}, which satisfies \eqref{eqn:xs_theorem2}. We complete the proof of the existence and the uniqueness of the equilibrium point of the mean-field model in \eqref{eqn:mean-field}. 

Next, we show the convergence to the equilibrium point $\bs{x}^*$ in \eqref{eqn:mean-field-solution}.
According to \cite[Theorem 3.2]{10.1145/3084454}, we need to show that $\bs{x}^*$ is locally exponentially stable and globally asymptotically stable (i.e., a fixed point to which all trajectories converge).
Particularly, by  \cite[Theorem 4.15]{Khalil2002}, we know $\bs{x}^*$ is (locally) exponentially stable for the dynamical system \eqref{eqn:mean-field}, if the corresponding linearized system at $\bs{x}^*$ is exponentially stable. 
Here, the Jacobian matrix of \eqref{eqn:mean-field} is given by
\begin{align}
\frac{\partial f}{\partial \bs{x}} =  \left( \begin{matrix} -\lambda & 0& \mu p \\
\lambda & -w(1-\gamma x_{\textit{S}}) & \mu (1-p) w\gamma x_{\textit{W}} \\
0 &w(1-\gamma x_{\textit{S}}) & -w\gamma x_{\textit{W}} -\mu \end{matrix} \right),
\end{align}
Let $\bs{q} \triangleq (q_{\textit{I}}, q_{\textit{W}}, q_{\textit{S}}) = \bs{x}- \bs{x}^*$. Then, the linearized system at $\bs{x}^*$ can be derived as
\begin{subequations}\label{eqn:linearized}
\begin{align}
        &\dot{q}_{\textit{I}} = -\lambda q_{\textit{I}} + \mu p q_{\textit{S}},\\
        &\dot{q}_{\textit{W}} =  \lambda q_{\textit{I}} -w(1-\gamma x_{\textit{S}}^*) q_{\textit{W}} + (\mu(1-p) +w\gamma x_{\textit{W}}^*) q_{\textit{S}},\\
        &\dot{q}_{\textit{S}} = w(1-\gamma x_{\textit{S}}^*) q_{\textit{W}} - (w\gamma x_{\textit{W}}^*+\mu) q_{\textit{S}}.
\end{align}
\end{subequations}
Now, by applying the Lyapunov method in a similar way to that for \cite[Theorem 2]{zhou2020meanfield}, we can prove that the linearized system in \eqref{eqn:linearized} is exponentially stable at $\bs{x}^*$. In particular, we introduce the Lyapunov function as $V(t) = |q_{\textit{I}}| + |q_{\textit{W}}|+|q_{\textit{S}}|$ and can show that 
\begin{align}\label{eqn:proof_thoerem2}
\dot{V}(t)\leq -\delta V(t),
\end{align}
where $\delta = \min\{\lambda,w(1-\gamma x_{\textit{S}}^*),w\gamma x_{\textit{W}}^* \}>0$.
This implies that 
$|q_{\textit{I}}| + |q_{\textit{W}}| + |q_{\textit{S}}| = V(t) \leq V(0)e^{-\delta t}.$
Thus, the linearized system in \eqref{eqn:linearized} is exponentially stable, implying that the dynamical system \eqref{eqn:mean-field} is locally exponentially stable at the equilibrium point $\bs{x}^*$. 
Moreover, from \eqref{eqn:proof_thoerem2}, we can see that the mean-field model in \eqref{eqn:mean-field} is globally asymptotically stable according to the Lyapunov theorem in \cite[Theorem 4.1]{Khalil2002}. 
Finally, by \cite[Theorem 3.2]{10.1145/3084454}, we can obtain the convergence properties of the mean-field model in \eqref{eqn:rates_of_convergence}, which completes the proof of Theorem~\ref{theorem:mean-field}.

\subsection{Proof of Lemma~\ref{lemma:properties_x}}\label{app:x}
We begin with policy (I). By \cite[Theorem 2]{zhou2020meanfield}, we know that the equilibrium point $\bs{x}^*$ under policy (I) satisfies:
\begin{subequations}\label{eqn:proof-properties-x-I}
\begin{align}
&x_{\textit{I}}^* = \frac{\mu}{\lambda}x_{\textit{S}}^*,\label{eqn:proof-xi-I}\\
&x_{\textit{W}}^* = \frac{\mu x_{\textit{S}}^*}{w(1-\gamma x_{\textit{S}}^*)},\label{eqn:proof-xw-I}\\
&x_{\textit{S}}^* = \frac{1/\mu}{1/\lambda+1/(w(1-\gamma x_{\textit{S}}^*))+1/\mu} = \frac{\lambda w (1-\gamma x_{\textit{S}}^*)}{(\lambda+\mu)w (1-\gamma x_{\textit{S}}^*) + \lambda\mu}.\label{eqn:proof-xs-I}
       \end{align}
\end{subequations}

\emph{1) Arrival rate $\lambda$:} 
By \eqref{eqn:proof-xs-I}, we can obtain the derivative of $x_{\textit{S}}^*$  with respect to $\lambda$:
\begin{align}
\frac{\partial x_{\textit{S}}^*}{\partial \lambda} = \frac{1/\mu}{(1/\lambda+1/(w(1-\gamma x_{\textit{S}}^*))+1/\mu)^2}\left(\frac{1}{\lambda^2} - \frac{\gamma}{w(1-\gamma x_{\textit{S}}^*)^2}\frac{\partial x_{\textit{S}}^*}{\partial \lambda}\right)
\end{align}
Thus, we have $\frac{\partial x_{\textit{S}}^*}{\partial \lambda}>0$. By \eqref{eqn:proof-xw-I}, we have
\begin{align}
&\frac{\partial x_{\textit{W}}^*}{\partial \lambda} = \frac{\mu}{w(1-\gamma x_{\textit{S}}^*)^2}\left(\frac{\partial x_{\textit{S}}^*}{\partial \lambda} (1-\gamma x_{\textit{S}}^*) + \gamma x_{\textit{S}}^*\frac{\partial x_{\textit{S}}^*}{\partial \lambda}\right)>0,\\
&\frac{\partial x_{\textit{I}}^*}{\partial \lambda}  = -\frac{\partial x_{\textit{W}}^*}{\partial \lambda} -\frac{\partial x_{\textit{S}}^*}{\partial \lambda}<0.
\end{align}

\emph{2) Service rate $\mu$:} 
By \eqref{eqn:proof-xs-I}, we can obtain the derivative of $x_{\textit{S}}^*$  with respect to $\mu$:
\begin{align}
\frac{\partial x_{\textit{S}}^*}{\partial \mu} = \frac{1}{((\lambda+\mu)w (1-\gamma x_{\textit{S}}^*) + \lambda\mu)^2} \left(-w\gamma \lambda^2 \mu \frac{\partial x_{\textit{S}}^*}{\partial \mu} - w(1-\gamma x_{\textit{S}}^*)\lambda(w(1-\gamma x_{\textit{S}}^*) +\lambda)\right)
\end{align}
Thus, we have $\frac{\partial x_{\textit{S}}^*}{\partial \mu}<0$.
By \eqref{eqn:proof-xi-I} and  \eqref{eqn:proof-xw-I}, we obtain:
\begin{align}
&\frac{\partial x_{\textit{I}}^*}{\partial \mu} = \frac{1}{\lambda} (x_{\textit{S}}^* + \mu \frac{\partial x_{\textit{S}}^*}{\partial \mu}),\label{eqn:proof-xi-mu}\\
&\frac{\partial x_{\textit{W}}^*}{\partial \mu} = \frac{1}{w (1-\gamma x_{\textit{S}}^*)^2}\left(x_{\textit{S}}^*(1-\gamma x_{\textit{S}}^*) + \mu \frac{\partial x_{\textit{S}}^*}{\partial \mu}\right).\label{eqn:proof-xw-mu}
\end{align}
As $x_{\textit{S}}^*(1-\gamma x_{\textit{S}}^*)\leq x_{\textit{S}}^*$, we can see that if $\frac{\partial x_{\textit{I}}^*}{\partial \mu}\leq 0$, then $\frac{\partial x_{\textit{W}}^*}{\partial \mu}\leq 0$, which contradicts to the observation that $\frac{\partial x_{\textit{I}}^*}{\partial \mu} + \frac{\partial x_{\textit{W}}^*}{\partial \mu}= - \frac{\partial x_{\textit{S}}^*}{\partial \mu} >0$. 
Thus, we must have $\frac{\partial x_{\textit{I}}^*}{\partial \mu} > 0$. Here, it is unknown whether $\frac{\partial x_{\textit{W}}^*}{\partial \mu}$ is positive or not.

\emph{3) Waiting rate $w$:} 
We calculate the derivative of $x_{\textit{S}}^*$ with respect to $w$ as follows:
\begin{align}
\frac{\partial x_{\textit{S}}^*}{\partial w} = \frac{1/\mu}{(1/\lambda+1/(w^2(1-\gamma x_{\textit{S}}^*))+1/\mu)^2}\left(1- \gamma x_{\textit{S}}^* -\gamma w \frac{\partial x_{\textit{S}}^*}{\partial w}\right),\label{eqn:rhs-xs-w}
\end{align}
which implies $\frac{\partial x_{\textit{S}}^*}{\partial w}>0$. By \eqref{eqn:proof-xi-I} and  \eqref{eqn:proof-xw-I}, we have:
\begin{align}
&\frac{\partial x_{\textit{I}}^*}{\partial w} = \frac{\mu}{\lambda}\frac{\partial x_{\textit{S}}^*}{\partial w}>0,~\frac{\partial x_{\textit{W}}^*}{\partial w}  = -\frac{\partial x_{\textit{I}}^*}{\partial \lambda} -\frac{\partial x_{\textit{S}}^*}{\partial \lambda}<0.
\end{align}

\emph{4) Successful transmission rate $p$:} 
From \eqref{eqn:proof-properties-x-I}, it can be seen that $\bs{x}^*$ is independent to $p$.

\emph{5) Ratio $\gamma$:} 
We calculate the derivative of $x_{\textit{S}}^*$ with respect to $\gamma$ as follows:
\begin{align}
\frac{\partial x_{\textit{S}}^*}{\partial \gamma} = -\frac{1/\mu}{(1/\lambda+1/(w(1-\gamma x_{\textit{S}}^*))+1/\mu)^2}\left(x_{\textit{S}}^* +\gamma\frac{\partial x_{\textit{S}}^*}{\partial \gamma}\right),\label{eqn:rhs-xs-gamma}
\end{align}
which implies $\frac{\partial x_{\textit{S}}^*}{\partial \gamma}<0$. By \eqref{eqn:proof-xi-I} and  \eqref{eqn:proof-xw-I}, we have:
\begin{align}
&\frac{\partial x_{\textit{I}}^*}{\partial w} = \frac{\mu}{\lambda}\frac{\partial x_{\textit{S}}^*}{\partial w}<0,~\frac{\partial x_{\textit{W}}^*}{\partial w}  = -\frac{\partial x_{\textit{I}}^*}{\partial \lambda} -\frac{\partial x_{\textit{S}}^*}{\partial \lambda}>0.
\end{align}

Similarly, we know that, under policy (W),  $\bs{x}^*$  satisfies \eqref{eqn:mean-field-solution-proof} and under policy (S), $\bs{x}^*$ satisfies:\begin{subequations}
\begin{align}
&x_{\textit{I}}^* = \frac{\mu p}{\lambda}x_{\textit{S}}^*,~x_{\textit{W}}^* = \frac{\mu p x_{\textit{S}}^*}{w(1-\gamma x_{\textit{S}}^*)},\label{eqn:proof-xixw-s}\\
&x_{\textit{S}}^* = \frac{1/\mu p}{1/\lambda+1/(w(1-\gamma x_{\textit{S}}^*))+1/\mu p} = \frac{\lambda w (1-\gamma x_{\textit{S}}^*)}{(\lambda+\mu p)w (1-\gamma x_{\textit{S}}^*) + \lambda\mu p}.\label{eqn:proof-xs-s}
\end{align}
\end{subequations}
By following the same approach for policy (I), we can  show the properties of $\bs{x}^*$ in terms of $\lambda$, $\mu$, $w$, and $\gamma$, and thus we omit the detailed calculations.
For $p$, under policy (W), we can obtain the corresponding derivative of $x_{\textit{S}}^*$ in \eqref{eqn:xs}:
\begin{subequations}
\begin{align}
\frac{\partial x_{\textit{S}}^*}{\partial p} = \frac{1/\mu p}{(p/\lambda+1/(w(1-\gamma x_{\textit{S}}^*))+1/\mu)^2}\left(-\frac{1}{\lambda} - \frac{\gamma}{w(1-\gamma x_{\textit{S}}^*)^2}\frac{\partial x_{\textit{S}}^*}{\partial p}\right),
\end{align}
\end{subequations}
which implies $\frac{\partial x_{\textit{S}}^*}{\partial p}<0$. Then, by \eqref{eqn:xixw}, we have
\begin{align}
\frac{\partial x_{\textit{W}}^*}{\partial p} = \frac{\mu}{w (1-\gamma x_{\textit{S}}^*)^2}\frac{\partial x_{\textit{S}}^*}{\partial p}<0,
~\frac{\partial x_{\textit{I}}^*}{\partial p} = -\frac{\partial x_{\textit{W}}^*}{\partial p} - \frac{\partial x_{\textit{S}}^*}{\partial p}>0.
\end{align}

Under policy (S), the derivative of $x_{\textit{S}}^*$ in \eqref{eqn:proof-xs-s} with respect to $p$ is obtained as follows:
\begin{align}
\frac{\partial x_{\textit{S}}^*}{\partial p} = -\frac{\lambda w}{((\lambda+\mu p)w (1-\gamma x_{\textit{S}}^*) + \lambda\mu p)^2}\left(\lambda\gamma\mu p \frac{\partial x_{\textit{S}}^*}{\partial p} +  (1-\gamma x_{\textit{S}}^*)(w\mu(1-\gamma x_{\textit{S}}^*) + \lambda\mu\right).
\end{align}
Thus, we have $\frac{\partial x_{\textit{S}}^*}{\partial p}<0$. By \eqref{eqn:proof-xixw-s}, we have 
\begin{align}
&\frac{\partial x_{\textit{I}}^*}{\partial p} = \frac{\mu}{\lambda} \left(x_{\textit{S}}^* + p \frac{\partial x_{\textit{S}}^*}{\partial p}\right),\label{eqn:proof-xi-mu-s}\\
&\frac{\partial x_{\textit{W}}^*}{\partial p} = \frac{\mu}{w (1-\gamma x_{\textit{S}}^*)^2}\left(x_{\textit{S}}^*(1-\gamma x_{\textit{S}}^*) + p \frac{\partial x_{\textit{S}}^*}{\partial p}\right).\label{eqn:proof-xw-mu-s}
\end{align}
As $x_{\textit{S}}^*(1-\gamma x_{\textit{S}}^*)\leq x_{\textit{S}}^*$, we can see that if $\frac{\partial x_{\textit{I}}^*}{\partial p}\leq 0$, then $\frac{\partial x_{\textit{W}}^*}{\partial p}\leq 0$, which contradicts to the observation that $\frac{\partial x_{\textit{I}}^*}{\partial p} + \frac{\partial x_{\textit{W}}^*}{\partial p}= - \frac{\partial x_{\textit{S}}^*}{\partial p} >0$. Thus, we must have $\frac{\partial x_{\textit{I}}^*}{\partial p}>0$. Here, it is unknown whether $\frac{\partial x_{\textit{W}}^*}{\partial p}$ is positive or not.
We complete the proof of Lemma~\ref{lemma:properties_x}.

\subsection{Proof of Theorem~\ref{theorem:properties-of-aoi}}\label{app:properties-of-aoi}
We first show that,  the average AoI under policies (I) and (S) as derived in Theorem~\ref{theorem:aoi} decreases with $k$. This can be obtained by taking the corresponding derivatives:
\begin{align*}
&\frac{\partial \bar{\Delta}_{\textrm{I-WP}}}{\partial k} = -\frac{\mu}{(\lambda+\mu)(\lambda+k)^2} -\frac{\lambda^2\mu^2+(1-p)k^2(\lambda^2+\mu^2)+\lambda\mu k^2(2-p)+2\lambda k\mu(\lambda+\mu)}{k^2(\lambda k + k\mu + \lambda\mu)^2}<0,\\
&\frac{\partial \bar{\Delta}_{\textrm{I-WOP}}}{\partial k}  = -\frac{1}{(\lambda+k)^2} -\frac{\lambda^2\mu^2+(1-p)k^2(\lambda^2+\mu^2)+\lambda\mu k^2(2-p)+2\lambda k\mu(\lambda+\mu)}{k^2(\lambda k + k\mu + \lambda\mu)^2}<0,\\
&\frac{\partial \bar{\Delta}_{\textrm{S-WP}}}{\partial k} = -\frac{\mu p}{(\lambda+\mu p)(\lambda+k)^2} -\frac{\lambda\mu p(k^2+2k\mu p+2\lambda k +\lambda\mu p)}{k^2(\lambda k + \mu p (\lambda+k))^2}<0,\\
&\frac{\partial \bar{\Delta}_{\textrm{S-WOP}}}{\partial k} = -\frac{1}{(\lambda+k)^2} -\frac{\lambda\mu p(k^2+2k\mu p+2\lambda k +\lambda\mu p)}{k^2(\lambda k + \mu p (\lambda+k))^2}<0.
\end{align*}
Similarly, we can show that the average AoI  in Theorem~\ref{theorem:aoi} decreases with $p$ and $\mu$. 

Next, based on Lemma~\ref{lemma:properties_x}, we show that, under policies (I) and (S), $k=w(1-\gamma x_{\textit{S}}^*)$ increases with $\mu$, $w$, and $p$, and decreases with $\gamma$. 
In particular, the monotonicity of $k$ with respect to $\mu$ and $p$ can be seen by noting that $\frac{\partial k}{\partial \mu} = - w\gamma \frac{\partial x_{\textit{S}}^*}{\partial \mu}>0$ and $\frac{\partial k}{\partial p} = - w\gamma \frac{\partial x_{\textit{S}}^*}{\partial p}>0$.
Moreover, according to \eqref{eqn:rhs-xs-w} and \eqref{eqn:rhs-xs-gamma}, we have $\frac{\partial k}{\partial w} = 1- \gamma x_{\textit{S}}^* -w\gamma \frac{\partial x_{\textit{S}}^*}{\partial w}>0$ and $\frac{\partial k}{\partial \gamma} = -w(x_{\textit{S}}^* + \gamma \frac{\partial x_{\textit{S}}^*}{\partial w})<0$.

Now, we can show that the monotonicity properties of the average AoI under policies (I) and (S) with schemes WP and WOP in the mean-field limit, in terms of $\mu$, $w$, $p$, and $\gamma$. 
From Theorem~\ref{theorem:aoi}, we can see that these expressions depend on $w$ and $\gamma$ only through $k=w(1-\gamma x_{\textit{S}}^*)$. 
As $\frac{\partial k}{\partial w}>0$, $\frac{\partial k}{\partial \gamma}<0$, and the expressions for policies (I) and (S) in Theorem~\ref{theorem:aoi} decrease with $k$, we immediately obtain that $\bar{\Delta}_{\textrm{I-WP}}(\bs{x}^*)$, $\bar{\Delta}_{\textrm{I-WOP}}(\bs{x}^*)$, $\bar{\Delta}_{\textrm{S-WP}}(\bs{x}^*)$, and $\bar{\Delta}_{\textrm{S-WOP}}(\bs{x}^*)$ decrease with $w$ and increase with $\gamma$.
To show the monotonicity in terms of $p$, we begin with policy (I) with scheme WP. With abuse of notation, let $k(p)=w(1-\gamma x_{\textit{S}}^*(p))$. If $p_1<p_2$, then we have
\begin{align}
&\bar{\Delta}_{\textrm{I-WP}}(k(p_1),p_1) -  \bar{\Delta}_{\textrm{I-WP}}(k(p_2),p_2)\nonumber\\
&=\bar{\Delta}_{\textrm{I-WP}}(k(p_1),p_1) - \bar{\Delta}_{\textrm{I-WP}}(k(p_2),p_1)+\bar{\Delta}_{\textrm{I-WP}}(k(p_2),p_1)-  \bar{\Delta}_{\textrm{I-WP}}(k(p_2),p_2)>0.
\end{align}
This is because $k(p_1)<k(p_2)$ and the expression of $\bar{\Delta}_{\textrm{I-WP}}$ in \eqref{eqn:avg_aoi_wp-I} decreases with $k$ and $p$.
Thus, $\bar{\Delta}_{\textrm{I-WP}}(\bs{x}^*)$ decreases with $p$. 
Similarly, we can show that, $\bar{\Delta}_{\textrm{I-WOP}}(\bs{x}^*)$, $\bar{\Delta}_{\textrm{S-WP}}(\bs{x}^*)$, and $\bar{\Delta}_{\textrm{S-WOP}}(\bs{x}^*)$ decrease with $\mu$. The monotonicity with respect to $\mu$ can also be shown in the same way. We complete the proof of Theorem~\ref{theorem:properties-of-aoi}.

\bibliographystyle{IEEEtran}
\bibliography{IEEEabrv,ref}

\end{document}